\documentclass[prb,twocolumn,superscriptaddress,aps,amsmath,amsfonts,notitlepage,longbibliography]{revtex4-2}
\usepackage{amsmath,amsfonts,amssymb,dsfont,graphicx,bm,caption,subcaption}
\usepackage{color}
\usepackage{hyperref}
\usepackage{tikz}
\usepackage{pbox}
\usepackage{balance}

\usepackage{txfonts}

\usepackage[normalem]{ulem}

\def\be{\begin{equation}}
\def\ee{\end{equation}}
\def\bea{\begin{eqnarray}}
\def\eea{\end{eqnarray}}
\def\bpm{\begin{pmatrix}}
\def\epm{\end{pmatrix}}

\def\Im{\mathop{\rm Im}}
\def\Re{\mathop{\rm Re}}

\def\Tr{\mathop{\rm Tr}}

\newcommand{\p}{\partial}

\begin{document}
\title{Tuneable band topology and optical conductivity in altermagnets}

\author{Peng Rao}
\affiliation{Physics Department, Technical University of Munich, TUM School of Natural Sciences, 85748 Garching, Germany}
\author{ Alexander Mook} 
\affiliation{Institute of Physics, Johannes Gutenberg University Mainz, Staudingerweg 7, Mainz 55128, Germany}
\author{Johannes Knolle}
\affiliation{Physics Department, Technical University of Munich, TUM School of Natural Sciences, 85748 Garching, Germany}
\affiliation{Munich Center for Quantum Science and Technology (MCQST), Schellingstr. 4, 80799 München, Germany}
\affiliation{Blackett Laboratory, Imperial College London, London SW7 2AZ, United Kingdom}

\date{\today}
\begin{abstract}
	We study two-dimensional $d$-wave altermagnetic metals taking into account the presence of substrate-induced Rashba spin-orbit coupling. We consider the altermagnet bandstructure using a 2D band Hamiltonian near the $\Gamma$ point under external magnetic field. It is shown that time-reversal-symmetry breaking due to altermagnetism, together with Rashba coupling and external magnetic field, can result in non-trivial band topology. The topological phases can be tuned by magnetic field strength and directions, and are classified by their Chern numbers. Furthermore, we investigate the charge response by computing the full optical conductivity tensor with and without magnetic field. In particular, we focus on magneto-optical responses, which are the finite-frequency analog of the Berry curvature-induced anomalous Hall conductivity. Finally, using experimentally realistic parameters for RuO$_2$, we estimate the Faraday angle in the absence of magnetic fields.
\end{abstract}

\maketitle

\section{Introduction}

Electronic systems without time-reversal-symmetry (TRS) and, hence, finite Berry curvature of the bandstructure can display intriguing (Hall) transport properties, most prominently the anomaous Hall effect (AHE) \cite{nagaosa2010anomalous}. The origin of the latter can be multifold depending on the interplay between topology of the electronic structure, different types of TRS breaking magnetic order and spin orbit coupling~\cite{xiao2010berry,chang2023colloquium}. A prominent example are magnetic Weyl semimetals~\cite{nagaosa2020transport} where the intrinsic Berry curvature can be probed by magneto-optical measurements. In general, the optical conductivity is not only a powerfull probe but also determines the frequency dependent permittivity and, thus, the coupling to electromagnetic waves. A strong off-diagonal component has potentially enormous practical implications for manufacturing quantum devices that utilise non-reciprocity, i.e. in quantum circulators~\cite{Viola2014,Mahoney2017,Bosco2017,Mahoney2017-2}. In many cases TRS is broken `extrinsically' by applied magnetic fields, which makes it difficult to implement these platforms in practical devices. Here new types of metallic magnets with intrinsically broken TRS and their heterostructures could lead to quantitative improvements.

\begin{figure}
	\centering
	\includegraphics[width=\linewidth]{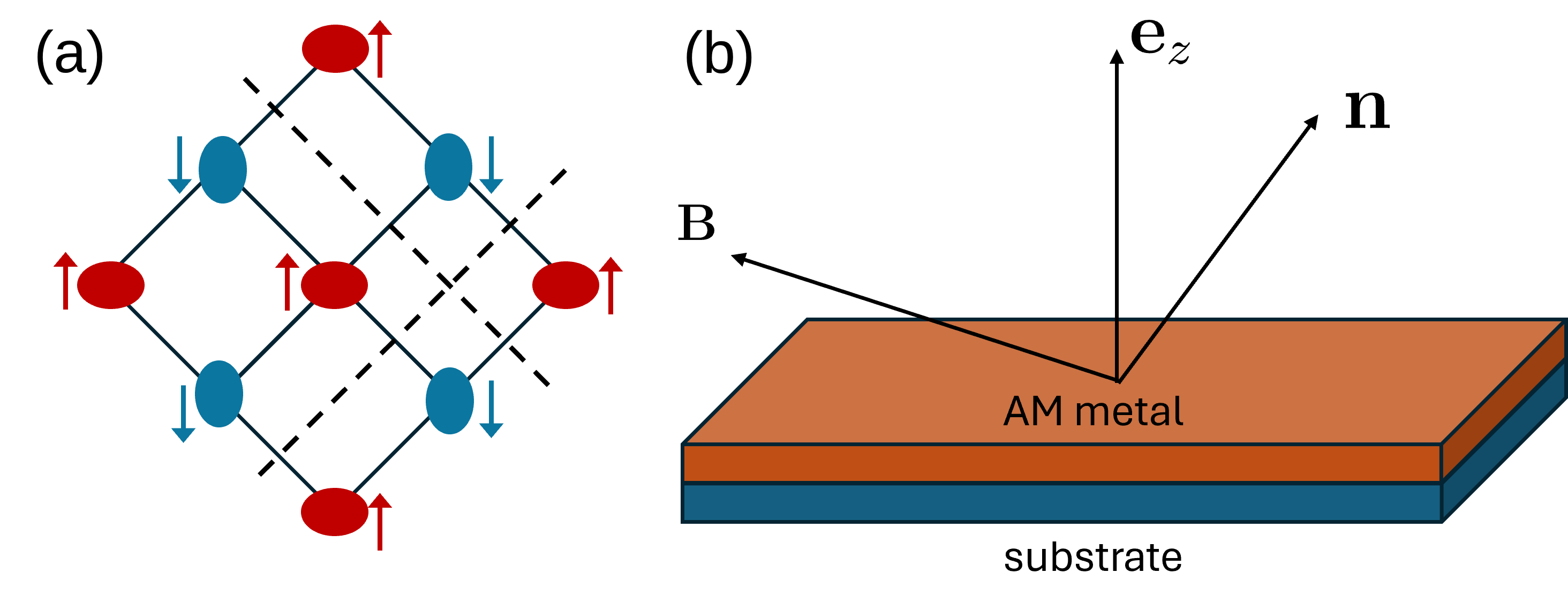}
	\caption{ Schematic plots of altermagnetism (AM) on a square lattice. (a) The magnetic order on the square lattice. The spin up (red) and down (blue) sublattices are shown explicitly. The deformation of magnetic atom orbitals is represented by red and blue ellipses. The resulting AFM order does not have combined translation or inversion and time-reversal symmetry $T$, but is symmetric under combined real space $C_4$ rotation and $T$. The system is additionally invariant under $T$ and real-space mirror reflections; the two mirror planes intersecting the lattice are shown as dashed lines. (b) 2D AM metal on a substrate, which induces interfacial spin-orbit coupling, and a N\'{e}el vector of arbitrary direction $\bm{n}$ determined by the intrinisc anisotropy and applied magnetic field $\bm{B}$.
	}
	\label{fig:AMSchematic}
\end{figure}

Altermagnetism (AM), as defined in Ref.~\onlinecite{Libor2022,Libor2022-1}, describes a newly discovered class of collinear magnetic materials that break TRS by a compensated order, but differ qualitatively from conventional collinear antiferromagnets (AFM) in their electronic properties~\cite{Hayami2019,Yuan2020,Libor2022,Libor2022-1}. In AMs, sublattices of opposite spin are not related by simple translation or inversion but require non-trivial rotations \cite{Libor2022,Libor2022-1}, for example because a low crystal environment leads to anisotropic \textit{spin densities} even though the absolute magnitude of the sublattice magnetization are equal [see Fig.~\ref{fig:AMSchematic}~(a)] or because of spontaneous orbital ordering \cite{Leeb2024orbitalorderingaltermagnetism}. Altermagnets then have only combined point group and TRS, and can be classified using the spin group formalism~\cite{Libor2022}. A key consequence -- and a reason for the large recent interest into AMs~\cite{Libor2022,Libor2022-1,Mazin2023Editorial, McClarty2024LandauTheory} -- is the strong splitting of spin-polarized electron bands on the altermagnetic background. Crucially, spin-splitting is not set by relativistic spin orbit coupling but can be large comparable to electronic energy scales. The unusual spin-split bandstructure from strong TRS breaking leads to unusual transverse responses of spin and charge, and magneto-optical responses in AMs \cite{Smejkal2020CrAHE, GonzalezHernandez2021,Ma2021,Zhou2021, Smejkal2022giant, Feng2022,Betancourt2023, Mazin2023, Krempasky2024, Fang2023, Tschirner2023, Zhou2024, Lee2024, Reimers2024, Fedchenko2024, Hariki2024}. 

In this paper, we present a comprehensive study of the non-trivial band topology and calculate the finite frequency optical response of AM thin films with substrate induced Rashba coupling, as depicted in Fig.~\ref{fig:AMSchematic} (b). Band topology in altermagnets is also investigated in Refs.~\cite{Fernandes2024,Antonenko2024}. We then study the associated optical Hall response which has recently also been discussed for the AM candidate FeSb$_2$~\cite{Attias2024,mazin2021prediction}. We show how spin-orbit coupling together with the strong TRS breaking spontaneously generates a nonzero Hall conductivity without external field, in agreement with Ref.~\cite{Smejkal2020CrAHE}. We analyze the resulting band topology by computing the Chern number, taking spin-orbit coupling strength, background N\'{e}el magnetisation direction $\bm{n}$, and the field $\bm{B}$ as parameters; rotating $\bm{n}$ can be achieved in principle by applying an external magnetic field~\footnote{The field magnitude required would depend on the strength of the easy-axis anisotropy. For RuO$_2$ the anisotropy along the $c$-axis is significant~\cite{Feng2022}, but weak magnetic anistropy might be present in other AM candidate materials.}. For $\bm{n}$ perpendicular to the surface, we identify two non-trivial topological phases characterized by the Chern number, which can be tuned by magnetic field directions. In particular, as the magnetic field rotates in-plane, the system undergoes four topological transitions in which the Chern number changes by unity. 

The TRS breaking of AMs results in an anti-symmetric Hall component of the optical conductivity tensor. We show that the Hall response at arbitrary frequency is expressed as a momentum integral over a function proportional to the Berry curvature, and tends to the anomalous Hall conductivity at vanishing frequency. The flexibility in engineering band topology mentioned above can then be used to tune the anomalous Hall conductivity. Note that the anomalous Hall conductivity is not related directly to the band Chern number, as non-trivial Berry curvature can generate a non-zero Hall conductivity even though the Chern number might still be zero~\cite{Smejkal2020CrAHE}. In particular, the anomalous Hall conductivity can be non-zero even without magnetic field. Finally, we estimate the Faraday effect from optical conductivity without magnetic field. Using microscopic band parameters, which have been discussed for the AM candidate RuO$_2$~\cite{Ko2018,Libor2022-1}, we estimate the order of magnitude of the Faraday angle to be: $\theta_F \sim 10^{-5}$~rad. 
Overall, TRS breaking responses can arise in AM in the absence of external magnetic field or net magnetization, which can be advantageous for applications, i.e. in quantum circulators.

The rest of the paper is organized as follows. Sec.~\ref{Sec:Model} introduces the low-energy Hamiltonian of a d-wave AM and discusses its electronic properties. There, we also study and map out the band topology in different parameter regimes including spin orbit coupling and external magnetic fields. In Sec.~\ref{Sec:Conductivity} 
we compute explicitly the optical conductivity as a function of frequency. In particular, we compute the anomalous Hall conductivity as a function of band parameters and relate the results to the band topology in the previous section. The results are then used to calculate the Faraday angle in Sec.~\ref{Sec:FaradayEffect}. We conclude in Sec.~\ref{Sec:Conclusion}

\section{Model and band topology}\label{Sec:Model}

We consider a minimal model of a two-dimensional (2D) d-wave metallic AM on a substrate. An effective low-energy model expanded to quadratic order in momenta around the $\Gamma$ point (Brillouin zone center) is given by the Hamiltonian~\cite{Smejkal2022}
\begin{equation}\label{eq:Hamiltonian}
H = \frac{\bm{p}^2}{2m}  +  \alpha \left(\bm{\sigma}\times \bm{p}\right) \cdot \bm{e}_z - \mu_B \bm{\sigma} \cdot \bm{B} + H_{\text{alt}}.
\end{equation}
Here, $m$ is the effective mass, $\alpha$ quantifies the Rashba spin-orbit coupling induced by the substrate, and $\bm{\sigma}$ is a vector of the spin Pauli matrices. The last term in Eq.~\eqref{eq:Hamiltonian} is due to the AM background, which has combined real space $4$-fold rotation and time-reversal symmetry $C_{4}T$:
\begin{equation}
H_{\text{alt}} = \frac{\beta_\text{M}}{2} \left(p_x^2 - p_y^2 \right) \bm{n} \cdot \bm{\sigma}.
\end{equation}
The unit vector $\bm{n}$ is parallel to the background N\'{e}el magnetisation. The strength of this coupling is quantified by $\beta_\text{M}$. The altermagnetic contribution is a non-relativistic effect that conserves the electron spin. D-wave magnetism can be read off from $H_{\text{alt}}$ by noting that it vanishes along the diagonals $p_x =\pm p_y$. Note that $\bm{n}$ changes sign under time-reversal. We assume that $\bm{n}$ is an independent parameter and neglect the coupling between $\bm{n}$ and the magnetic field. We also neglect orbital effects from the perpendicular component of $\bm{B}$. The model (\ref{eq:Hamiltonian}) is represented schematically in Fig.~\ref{fig:AMSchematic}~(b).

It is convenient to rewrite Eq.~(\ref{eq:Hamiltonian}) as
\begin{equation}\label{eq:Hamiltonian-1}
H = \frac{\bm{p}^2}{2m} +\bm{N}(\bm{p}) \cdot \bm{\sigma},
\end{equation}
with
\begin{equation}\label{eq:Omega}
\bm{N}(\bm{p})= \frac{\beta_M}{2} \left(p_x^2 - p_y^2 \right) \bm{n}+ \alpha \bm{p}\times \bm{e}_z - \mu_B \bm{B}. 
\end{equation}
The eigenvalues of the Hamiltonian \eqref{eq:Hamiltonian-1} are compactly expressed as
\begin{equation}\label{eq:Eigenvalues}
\varepsilon_{\pm}(\bm{p}) = \frac{p^2}{2m} \pm |\bm{N}(\bm{p})|,
\end{equation}
and the respective eigenstates $\eta_\pm(\bm{p})$ are spinors with spin polarized along $\pm \bm{N}$. 

\begin{figure}
	\centering
	\includegraphics[width=\linewidth]{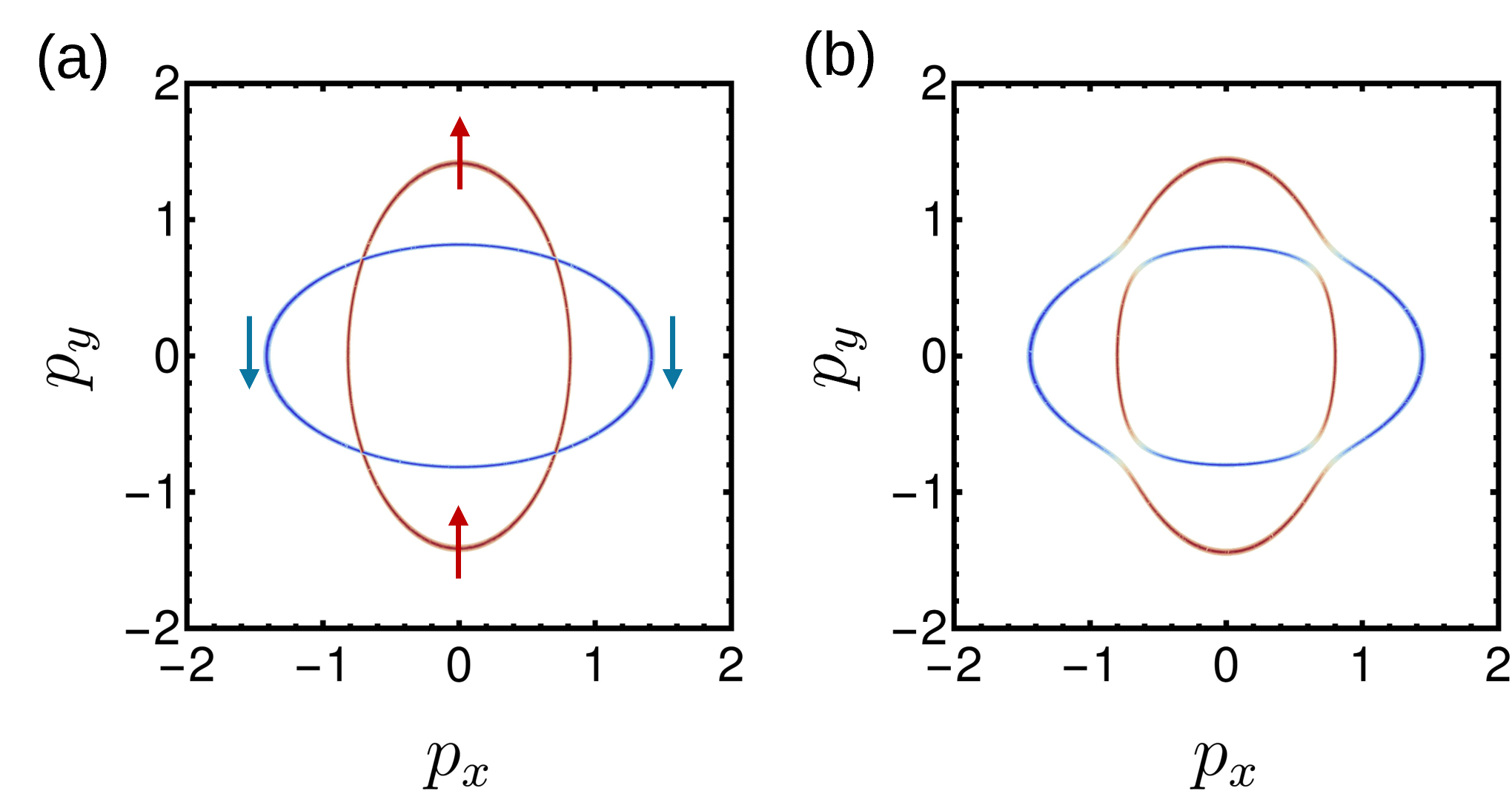}
	
	\caption{Fermi surfaces at $\bm{n}=(0,0,1)^\text{T}$ and $\mu=0.5 t$ (a) without spin-orbit coupling; and (b) $\alpha = 0.1 t$. The electron $S_z$ spin component along the Fermi surface is shown in color: red and blue are spin up and down respectively. Spin-orbit coupling couples spin to momentum, and splits the elliptical Fermi surfaces at the four points along the diagonals, where the $S_z$ components vanish. The Fermi surfaces show the $C_4T$ symmetry and the mirror symmetries along two vertical planes that intersect the two BZ diagonals. 
	}
	\label{fig:Spectrum}
\end{figure}

First we shall discuss the effect of each term in Hamiltonian \eqref{eq:Hamiltonian-1} on the electronic band structure. Without spin-orbit coupling and magnetic field, i.e., $\alpha = |\bm{B}| = 0$, the Fermi surface at positive chemical potential $\mu$ consists of two identical ellipses at an angle $\pi/2$ with respect to each other and intersecting at the diagonals $p_x=\pm p_y$, with spins polarised along $\pm \bm{n}$ [see Fig.~\ref{fig:Spectrum}(a)]. The ellipses' major and minor axes $a$ and $b$ are given by
\begin{subequations}
	\begin{align}
	a^2 &= \frac{2\mu}{(1/m) + \beta_M}, \\
	b^2 &= \frac{2\mu}{(1/m) - \beta_M}.
	\end{align}
\end{subequations}

For physical systems, $\mu_B |\bm{B}|, \alpha p_F \ll \beta_M p_F^2\sim \mu$ where $p_F= 2m\mu$ is the Fermi momentum at the intersections. Therefore, the inclusion of a small spin-orbit coupling splits the Fermi-surfaces at the four intersection points by $\delta p_\perp \sim 2 \alpha p_F /v_F \ll p_F$ where $v_F = p_F/m$ is the Fermi velocity at these points [see Fig.~\ref{fig:Spectrum})(b)]. For $\bm{n}$ along the $z$-axis without magnetic field, the Hamiltonian \eqref{eq:Hamiltonian-1} has magnetic group $C_{4}'$ symmetry, and two additional magnetic point group mirror planes along the diagonals of the Brillouin zone (BZ). The shape of the split Fermi surfaces retains the $C_4$ symmetry. Similarly, the magnetic field only distorts slightly the Fermi surface. For both $\bm{B}, \bm{n}$ along the $z$-axis, the shape of the Fermi-surfaces is $C_4$ symmetric. However, the magnetic field additionally induces a gap of $ 2 \mu_B |\bm{B}|$ at the origin, where the bands are otherwise degenerate. 

In the rest of this paper, we shall take the lattice constant to be unity. Thus momentum $\bm{p}$ is dimensionless, whereas $1/m, \alpha$ and $\beta_M$ have the units of energy. We shall take as the unit of energy $t$ which is of the order of the band-width. A typical value is $t\sim 1$~eV. For numerical simulations below, we shall express all band parameters in the reduced unit and set $1/m=t, \beta_M =\mu =  0.5t, p_F=1$ and $ \alpha p_F = 0.1 t$, unless otherwise specified. Frequency $\omega$ is also expressed in units of $t$.

We now discuss the band topology of our system, which originates from the spin-structure of the Hamiltonian \eqref{eq:Hamiltonian-1} in momentum space. The band topology is determined by the the Berry curvature of the lower band
\begin{equation}\label{eq:BerryCurvature}
\Omega(\bm{p}) = \frac{1}{2} \widehat{\bm{N}} \cdot \left(\frac{\p \widehat{\bm{N}}}{\p p_x} \times \frac{\p \widehat{\bm{N}}}{\p p_y}\right), \
\quad \widehat{\bm{N}} = \frac{\bm{N}}{|\bm{N}|},
\end{equation}
and the Chern number
\begin{equation}\label{eq:ChernNumber}
C =  \frac{1}{2\pi}\int \Omega(\bm{p})  d^2p.
\end{equation}
Thus, the topology of the system is determined by the interplay between Rashba spin-orbit coupling $\alpha$, the vector $\bm{n}$, and the external field $\bm{B}$, all of which which enter in $\bm{N}$ in Eq.~\eqref{eq:Omega}. In the rest of this section, we shall discuss first the case without magnetic field, where the system is gapless and $C$ is ill-defined. We then move to three gapped cases due to external field, which exhibit non-trivial band topology.

For later convenience, we derive another expression for $\Omega(\bm{p}) $. By substituting the differential identity
\begin{align}
d \widehat{\bm{N}} = \frac{d\bm{N}}{|\bm{N}|} - \frac{\widehat{\bm{N}}}{|\bm{N}|} d |\bm{N}|
\end{align}
into Eq.~\eqref{eq:BerryCurvature}, we obtain
\begin{equation}\label{eq:BerryCurvature-1}
\Omega(\bm{p}) = \frac{1}{2|\bm{N}|^2} \widehat{\bm{N}} \cdot \left(\frac{\p \bm{N}}{\p p_x} \times \frac{\p \bm{N}}{\p p_y}\right).
\end{equation}

\begin{figure}
    \centering
    \includegraphics[width=\linewidth]{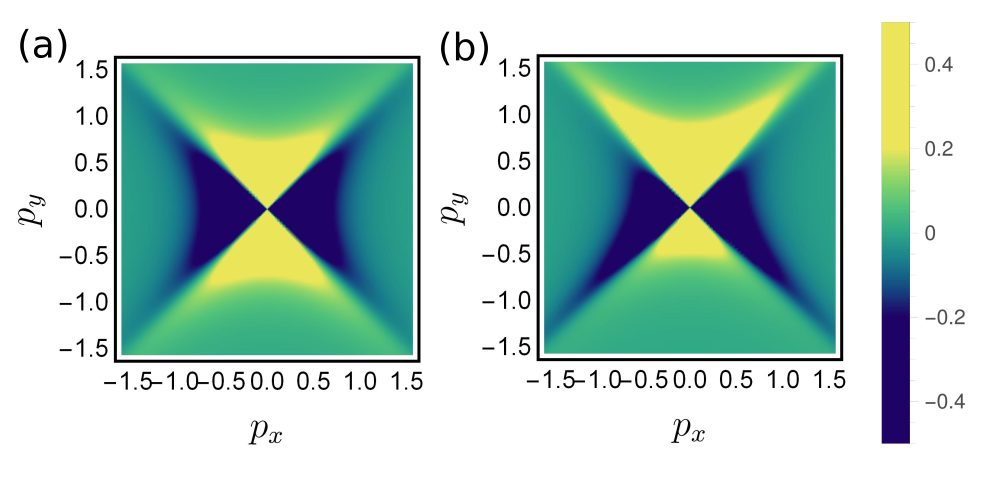}
    \caption{Berry curvature $\Omega(\bm{p}) $ as a function of momentum for selected $\bm{n}$ directions at $\bm{B}=0$: (a) $\Omega(\bm{p}) $ for $\bm{n}=(0,0,1)^\text{T}$; (b) $\Omega(\bm{p}) $ for $\bm{n}=(1,0,1)^\text{T}/\sqrt{2}$. Note the gap closing at the origin which appears as a singularity in $\Omega(\bm{p}) $.
    }
    \label{fig:ChernNumber(H=0)}
\end{figure}

First, we consider the case of zero magnetic field, $\bm{B}=0$. The two bands touch at the origin, and the Chern number is ill-defined. Near the origin, the Berry curvature exhibits d-wave structure with large quadrupole moments. For $\bm{n}$ along the $z$-axis, the Berry curvature reads~\cite{Smejkal2022} 
\begin{equation}
\Omega(\bm{p})  = - \frac{\alpha \beta_M^2(p_x^2-p_y^2)}{2\left[\alpha^2p^2+ \beta_M^2(p_x^2-p_y^2)^2/4\right]^{3/2}}.
\end{equation}
the dipole moments compensate each other to give $C = 0$ [see Fig.~\ref{fig:ChernNumber(H=0)}(a)]. At $\bm{n}$ away from the $z$-axis, the quadrupoles are generally not perfectly compensated [see Fig.~\ref{fig:ChernNumber(H=0)}(b)].

The external magnetic field gaps out the bands and the Chern number becomes well-defined. In the rest of this section, we take $\bm{n}= \bm{e}_z$ and study the relation of the resulting topology to field directions. Note that spin-orbit coupling is still essential for band topology. In the absence of spin-orbit coupling, $\widehat{\bm{N}}$ lies entirely in the plane extended by the vectors $\bm{n}$ and $\bm{B}$, thus $\Omega(\bm{p}) $ given by Eq.~\eqref{eq:BerryCurvature} is identically zero.

\subsection{Perpendicular magnetic field}

\begin{figure}
    \centering
    \includegraphics[width=\linewidth]{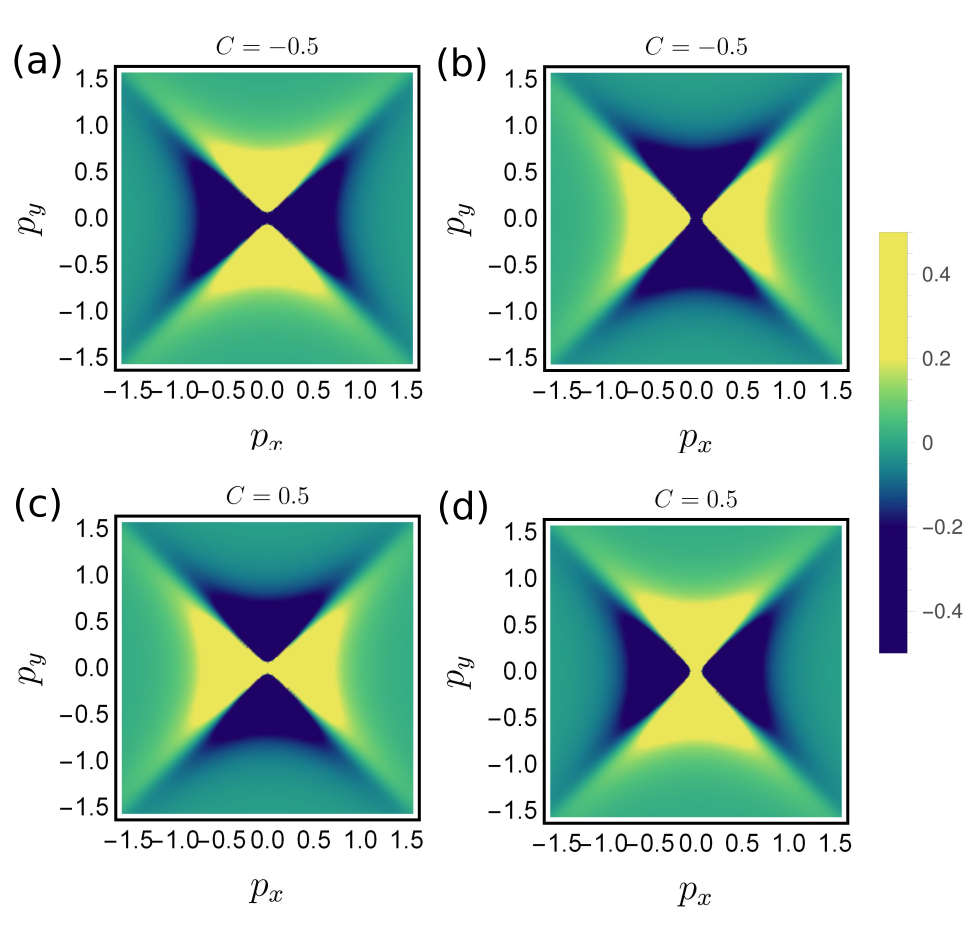}
    \caption{Berry curvature $\Omega(\bm{p}) $ as a function of momentum for $\bm{n}=(0,0,1)^\text{T}$. Upper column: $\bm{B}=0.01(0,0,1)^\text{T}$ for $\beta_M = 0.5$ (left) and $\beta_M = -0.5t$ (right). Lower column: $\bm{B}=-0.01(0,0,1)^\text{T}$ for $\beta_M = 0.5t$ (left) and $\beta_M = -0.5t$ (right). Comparing to Fig.~\ref{fig:ChernNumber(H=0)}, the origin is no longer singular due to gap opening from magnetic field. In all cases the Chern number $C$ does not depend on $\beta_M$ in agreement with Eq.~(\ref{eq:ChernNumber-PerpH}). Note the large absolute values of $\Omega(\bm{p}) $ near the origin, the sign being determined by $|\bm{B}|$.
    }
    \label{fig:ChernNumberHperpendicular}
\end{figure}

We consider a magnetic field along the $z$-axis, that is, $\bm{B} = (0,0,B)^\text{T}$. It is sufficient to consider a small field, because the band topology at larger field values is the same. The dominant contributions to the Chern number $C$ in Eq.~\eqref{eq:ChernNumber} come from around the origin. Therefore, it is sufficient to consider $\Omega(\bm{p}) $ at small $\bm{p}$ where the altermagnetic term can be neglected. We find near the origin that
\begin{equation}
    \bm{N} \approx \left(\alpha p_y,-\alpha p_x, -\mu_B B\right)^\text{T},
\end{equation}
which is formally identical to the Rashba case where the magnetic field $B$ is equivalent to the gap term~\cite{nagaosa2010anomalous}. The Chern number reads~\cite{nagaosa2010anomalous}
\begin{equation}\label{eq:ChernNumber-PerpH}
    C = - \frac{1}{2} \text{sign}\ B. 
\end{equation}
Note that $C$ does not depend on $\beta_M$. We numerically confirm the conclusions above and show the results in Fig.~\ref{fig:ChernNumberHperpendicular}. Note that here the Chern number is a half-integer, since we have only considered the long-wavelength limit near the $\Gamma$ point. It is expected that AM will induce non-trivial topology at other regions of the BZ where the bands are gapped. The total Chern number must add up to an integer.

\subsection{In-plane magnetic field}
\label{sec:inplanefield}
An in-plane magnetic field, $\bm{B} = (B_x, B_y, 0)^\text{T}$, induces tunable band topology, as can be seen from 
\begin{equation}
    \bm{N} = \left(\alpha p_y -\mu_B B_x, -\alpha p_x-\mu_B B_y, \frac{\beta_M}{2}(p_x^2-p_y^2) \right)^\text{T}.
\end{equation}
The magnetic field shifts the center of the Berry curvature quadruple moment from the origin to momentum
\begin{equation}\label{eq:MomentumOrigin}
\bm{p}_0 = \frac{\mu_B}{\alpha}(-B_y,B_x,0)^\text{T},
\end{equation}
see Figs.~\ref{fig:ChernNumberHinPlane}(a,b). Analogous to the previous subsection, we take the external field to be small and shift the momentum origin to $ \bm{p}_0$. The vector $\bm{N}$ near $\bm{p}_0$ now reads
\begin{equation}
    \bm{N} \approx \left(\alpha p_y , -\alpha p_x, \frac{\beta_M}{2}(p_{0,x}^2-p_{0,y}^2) \right)^\text{T}.
\end{equation}
Although it is again formally identical to the Rashba case, the gap is now provided by the altermagnet term. As a result, the Chern number is given by 
\begin{equation}\label{eq:ChernNumber-InPlaneH}
    C = \frac{1}{2}\text{sign}\left[\beta_M(p_{0,x}^2-p_{0,y}^2)\right],
\end{equation} 
and can be tuned by in-plane field directions. In particular, the gap closes at $p_{0,x} = \pm p_{0,y}$ corresponding to $B_x= \pm B_y$, and the system undergoes a series of topological transitions as $\bm{B}$ is rotated in the plane. These topological transitions are shown in Fig.~\ref{fig:ChernNumberHinPlane}(c) for $\bm{B} = B ( \cos \phi, \sin \phi, 0)^\text{T}$ and $\phi \in (0,2\pi)$. Note that the gap closing occurs when $\bm{B}$ lies within the mirror planes. The topological transitions described above are similar to a case in the context of Weyl semi-metals considered in Ref.~\cite{You2019}.

We note in passing that if $\bm{n}$ is also rotated in-plane, $\Omega(\bm{p}) $ vanishes identically. This is because $\widehat{\bm{N}}$ now lies entirely within the $xy$-plane, and Eq.~(\ref{eq:BerryCurvature}) is identically zero.

\subsection{Arbitrary magnetic field directions}
\begin{figure}
    \centering
    \includegraphics[width=\linewidth]{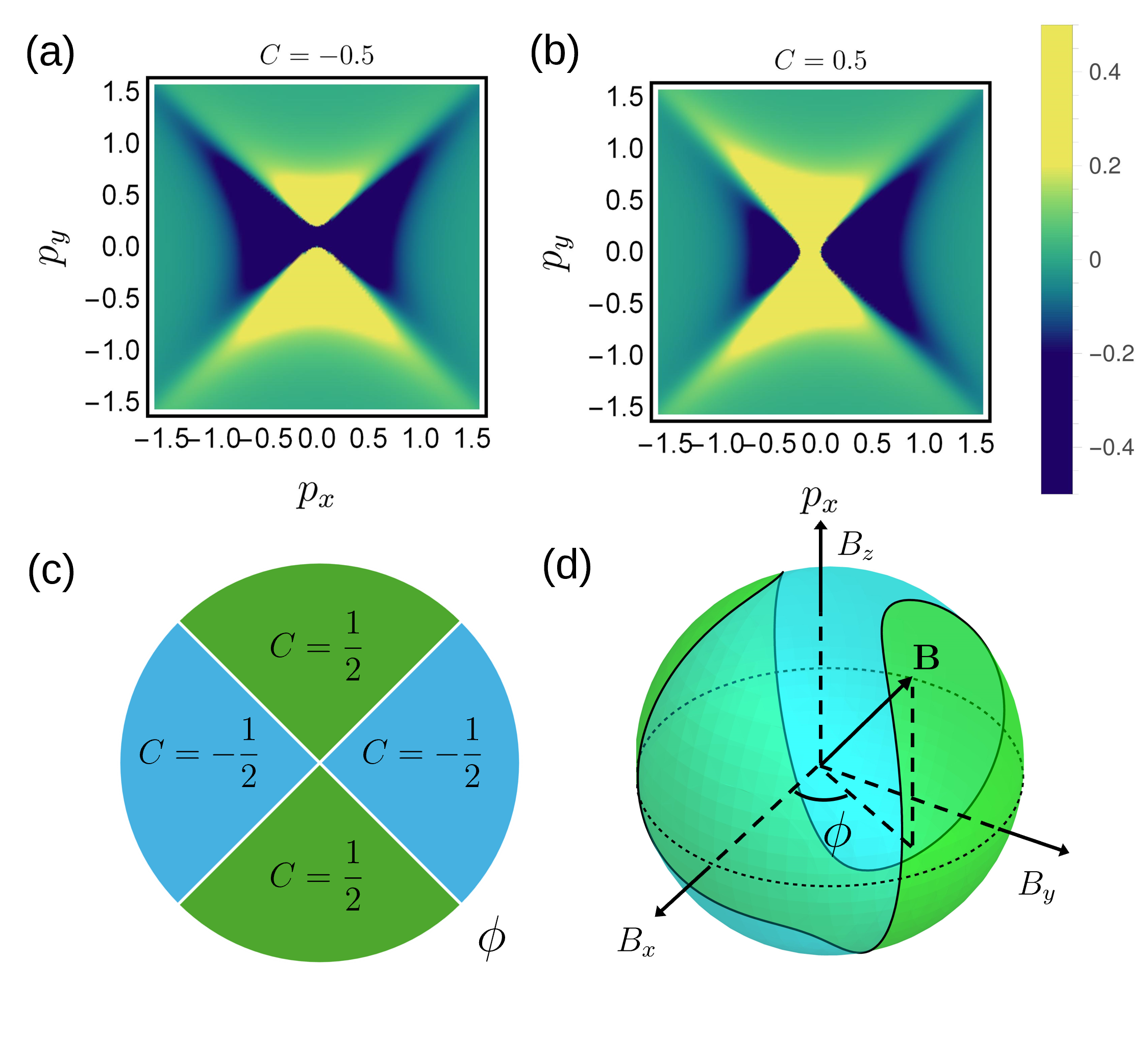}
    \caption{Berry curvature and band topology for $\bm{n}=(0,0,1)^\text{T}$ and in-plane magnetic field. Upper row: Berry curvature $\Omega(\bm{p}) $ as a function of momentum at (a) $\bm{B}=0.01(1,0,0)^\text{T}$; and (b) $\bm{B}=0.01(0,1,0)^\text{T}$. The Chern numbers $C$ have the opposite sign in agreement with Eq.~(\ref{eq:ChernNumber-InPlaneH}). The center of the Berry curvature quadruple moment is shifted by $\bm{p}_0$ from the origin given by Eq.~(\ref{eq:MomentumOrigin}). Lower row:~(c) schematic plot of $C$ as a function of magnetic field polar angle $\phi$;~(d) Schematic phase diagram as a function of $\bm{B}$ directions given by Eq.~(\ref{eq:CriticalLine}). Blue and Green regions correspond to $C=\pm 1/2$ respectively. The dashed line is the equator which contains the phase diagram in (c). 
    }
    \label{fig:ChernNumberHinPlane}
\end{figure}

We now let $\bm{B}$ have arbitrary directions and constant magnitude. Analogous to before, the in-plane components of $\bm{B}$ shift the center of the Berry curvature dipole to $\bm{p}_0$, near which the $z$-component of $\bm{N}$ determines the gap:
\begin{equation}
    \bm{N} \approx \left(\alpha p_y , -\alpha p_x, \frac{\beta_M}{2}(p_{0,x}^2-p_{0,y}^2) - \mu_B B_z \right)^\text{T}.
\end{equation}
Thus, substituting Eq.~(\ref{eq:MomentumOrigin}) in $N_z$ above, the gap closing is determined by
\begin{equation}\label{eq:CriticalLine}
    B_x^2+B_y^2+B_z^2 = | \bm{B} |^2, \qquad \ B_z = \frac{\mu_B\beta_M}{2\alpha^2}(B_y^2-B_x^2). 
\end{equation}
Graphically Eq.~\eqref{eq:CriticalLine} means that the critical line in the space of $\bm{B}$ is the intersection of a hyperbolic paraboloid with a sphere; see Fig.~\ref{fig:ChernNumberHinPlane}(d). We see that there are two global topological phases. For $B_z = 0$, we re-encounter the four phase transitions discussed in Sec.~\ref{sec:inplanefield}.

Finally, note that there exists also two global topological phases with opposite $C=\pm 1/2$ for a system with Rashba SOC but without the AM term $H_{\text{alt}}$. There, the critical line is the equator in external field directions. As can be seen from Fig.~\ref{fig:ChernNumberHinPlane}(d), $H_{\text{alt}}$ pushes the critical line into the two hemispheres, which makes the in-plane topological transitions possible.

\section{Optical conductivity}\label{Sec:Conductivity}

We now discuss the linear electric conductivity in the system described by the Hamiltonian \eqref{eq:Hamiltonian-1}. First, let us discuss the general properties of conductivity under time-reversal transformation. The conductivity tensor $\sigma_{\alpha \beta}$ satisfies the Onsager relation,
\begin{equation}\label{eq:Conductivity-Symmetry}
    \sigma_{\alpha \beta}(\omega; \bm{n}, \bm{B})  =  \sigma_{\beta \alpha}(\omega; -\bm{n}, -\bm{B}),
\end{equation}
due to time-reversal operation. The time-reversal-odd quantities $\bm{n}$ and $\bm{B}$ change sign. Thus in a time-reversal-invariant system, $\sigma_{\alpha \beta}$ is symmetric. If time-reversal-symmetry is broken, the conductivity can be decomposed into symmetric $\sigma^{\text{s}}$ and anti-symmetric $\sigma^{\text{as}}$ components:
\begin{equation}
    \sigma_{\alpha\beta} = \sigma^{\text{s}}_{\alpha\beta}+ \sigma^{\text{as}}_{\alpha\beta}.
\end{equation}
Eq.~\eqref{eq:Conductivity-Symmetry} then gives that $\sigma^{\text{as}}$ ($\sigma^{\text{s}}$) is an odd (even) function of $\bm{n}$ and $\bm{B}$.

\begin{figure}
    \centering
    \includegraphics[width=\linewidth]{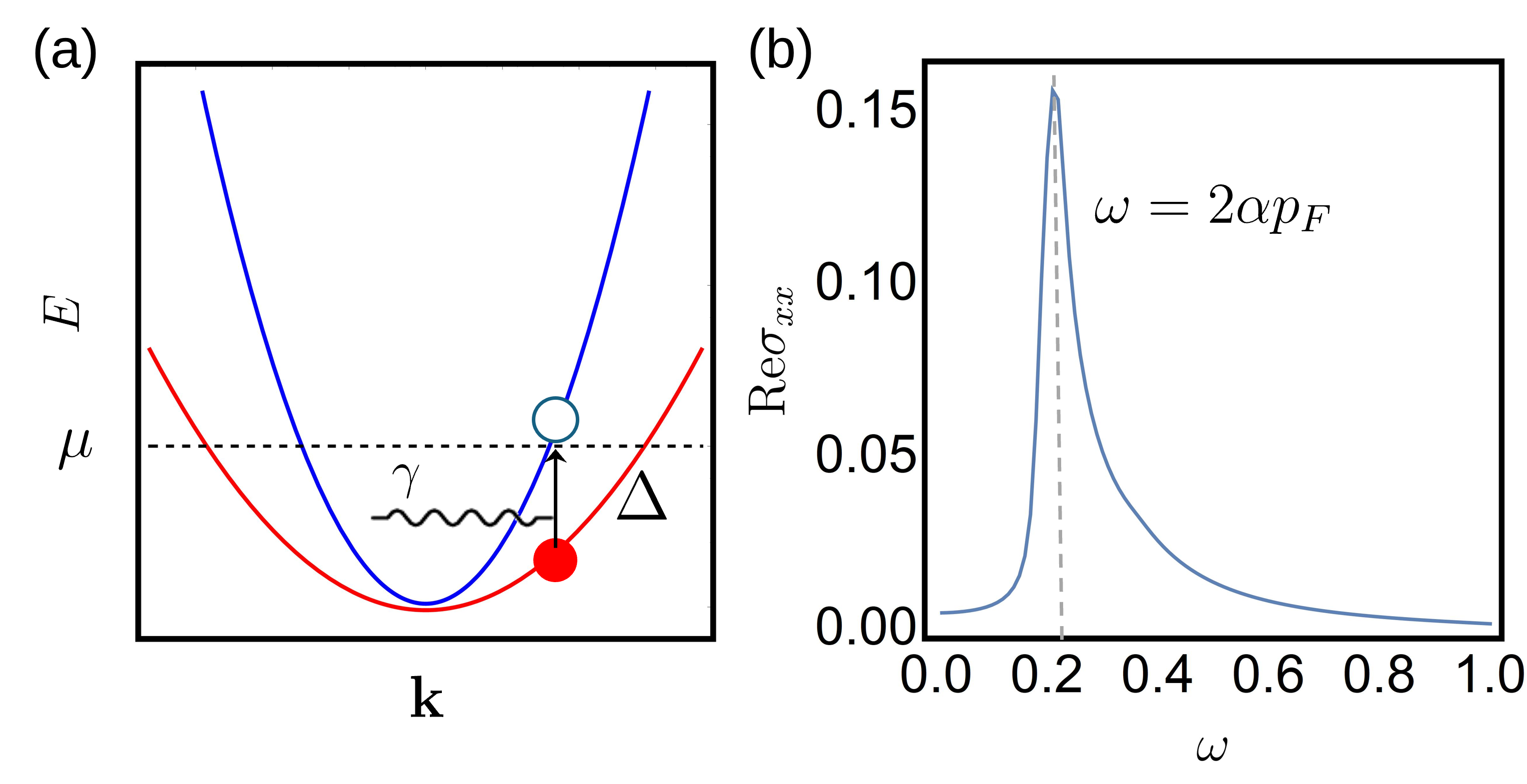}%
    \caption{ Real longitudinal conductivity due to inter-band transitions. (a) Schematic plot for absorption by inter-band transition that excites an electron from the filled lower band (red) to the unfilled upper band (blue). (b) Conductivity $\Re \sigma_{xx}$ at $\bm{n}=(0,0,1)^\text{T}$. The peak is at $\omega = 2\alpha p_F$, the Rashba splitting of Fermi surfaces. 
    }
    \label{fig:Absorption}
\end{figure}

There are additional constraints on $\sigma_{\alpha \beta}$ when dissipation can be neglected. In our system, this corresponds to frequencies $\omega $ below the minimum direct band-gap $\Delta$ between the two Fermi surfaces, which forbids direct inter-band transition from the filled lower band to the unfilled upper band; see Fig.~\ref{fig:Absorption}~(a). Here we also neglect dissipation due to impurity scattering. The additional constraint on $\sigma_{\alpha\beta}$ can be derived by setting to zero the energy dissipation per unit time: 
\begin{equation}
 - \text{Re} \ \bm{j} \cdot \text{Re} \ \bm{E} = -\frac{1}{4} \left( \bm{j} + \bm{j}^*\right) \cdot \left( \bm{E} + \bm{E}^*\right)=0.
\end{equation}
(This derivation is similar to that given in Ref.~\cite{Landau8} p.332, for the dielectric tensor $\varepsilon_{\alpha\beta}$.)
After substituting $j_\alpha =\sigma_{\alpha\beta}E_\beta$ and time averaging, the terms $E_\alpha^* E_\beta^*$ and $E_\alpha E_\beta$ vanish since they are proportional to $\exp(\pm 2i \omega t)$. We then have
\begin{equation}
 -\frac{1}{2}\sum_{\alpha\beta}\left( \sigma_{\alpha \beta} + \sigma_{ \beta \alpha}^*\right) E_\alpha E_\beta^*=0.
\end{equation} 
The conductivity is as a result anti-Hermitian in the absence of dissipation, that is to say
\begin{equation}\label{eq:NoDissipation}
   \sigma_{\alpha \beta}(\omega; \bm{n}, \bm{B}) = - \sigma_{ \beta \alpha}^*(\omega; \bm{n}, \bm{B}).
\end{equation}
Equations~(\ref{eq:Conductivity-Symmetry}) and (\ref{eq:NoDissipation}) mean that, at $\omega < \Delta$, $\sigma^{\text{s}}$ is imaginary whereas $\sigma^{\text{as}}$ is real and can be written as
\begin{equation}\label{eq:ASConductivity}
    \sigma^{\text{as}}_{\alpha\beta} = \varepsilon_{\alpha\beta \gamma} g_\gamma,
\end{equation}
where $\bm{g}$ is a real vector function that changes sign under time-reversal: $\bm{g}(\omega; \bm{n}, \bm{B}) = -\bm{g}(\omega; -\bm{n}, -\bm{B})$. In particular at $\omega \rightarrow 0$, $\sigma^{\text{as}}_{xy}$ tends to the anomalous Hall conductivity $\sigma_{\text{AH}}$ which results from non-trivial band Berry curvature of the system [see Eq.~(\ref{eq:AnomalousHallConductivity}) below].

We now compute the optical conductivity of the system (\ref{eq:Hamiltonian-1}) and study its dependence on the parameters $\bm{n}$ and $\bm{B}$. We set temperature $T=0.01 t \ll \alpha p_F$ for numerically convenience, which agrees with the $T=0$ limit since in all cases that we consider below, the energy gap $\Delta$ between the Fermi surfaces obeys $\Delta \gg T$. 

In what follows, we use the Mastubara Green's functions. The free electron Green's functions is written as
\begin{align}
     G_{\pm}(p) &= -\int_0^{1/T} d\tau  e^{ip_0 \tau} \langle \text{T}_\tau \{ a_\pm(\tau,\bm{p})  a_\pm^\dagger(0,\bm{p})\}\rangle  \\ 
    &= \frac{1}{i p_0 - \varepsilon_{\pm}(\bm{p})+\mu}.
\end{align}
Here $a_\pm (\tau, \bm{p})$ is the electron field of the $\pm$ band at imaginary time $\tau$, and T$_{\tau}$ denotes time-ordering. We have used four-vector notation $p =(p_0, \bm{p})$, where $p_0$ is the fermion Matsubara frequency. The free Green's function in spin-basis is given by
\begin{equation}\label{eq:GreensFunction}
    G (p) = \frac{1}{2}\left\{ G_+(p) + G_-(p) + \bm{\sigma} \cdot \widehat{\bm{N}} \left[ G_+(p) - G_-(p)\right]\right\}.
\end{equation}
Equation~(\ref{eq:GreensFunction}) can be derived by operating $G(p)$ on the eigenstates $\eta_\pm (\bm{p})$ which correspond to spin parallel and anti-parallel to $\widehat{\bm{N}}$. 

Futhermore, the electric current operator is defined as
\begin{equation}\label{eq:Current}
\begin{split}
 j_\alpha(\bm{k})   = - \frac{\p H(\bm{k}- e \bm{A})}{\p A_\alpha} = e\left[j_\alpha^{(0)}(\bm{k})+j_\alpha^{(1)}(\bm{k})\right],
\end{split}
\end{equation}
where $e$ is the electron charge, $\bm{A}$ is the vector potential and $\bm{j}^{(0)}$ and $\bm{j}^{(1)}$ are
\begin{align}
&j_\alpha^{(0)}(\bm{k}) = \frac{\p H(\bm{k})}{\p k_\alpha}, \\
&j_\alpha^{(1)}(\bm{k}) =- e\left( \frac{1}{m}A_\alpha + \beta_M \tau^z_{\alpha \beta} \bm{n}.\bm{\sigma} A_\beta \right).
\end{align}
Here $\tau^z$ is the third Pauli matrix. Note that $j_\alpha^{(1)}(\bm{k})$ is already linear in the perturbation. We choose the Coulomb gauge, $\text{div}\ \bm{A} = 0$, and set the scalar potential to zero. The conductivity is then evaluated as
\begin{align}\label{eq:Conductivity}
     \sigma_{\alpha \beta} &= \frac{ie^2}{\omega} \left[ \Pi_{\alpha \beta}(\omega)- \Pi_{\alpha \beta}(0)  \right] + \sigma^{(1)}_{\alpha \beta}.
\end{align}
Here, the first contribution is the Kubo response associated with the perturbed density matrix, and
\begin{equation}
     \sigma^{(1)}_{\alpha \beta} = i \frac{ e^2 }{\omega} \left(\frac{N_e}{m}\delta_{\alpha \beta} -\beta_M \tau^z_{\alpha \beta} \overline{\bm{n} \cdot \bm{\sigma}}\right),
\end{equation}
($N_e$ is the electron density) comes from directly averaging $\bm{j}^{(1)}$ over the equilibrium distribution (as indicated by the overline). This gives
\begin{equation}
\begin{split}
\overline{\bm{n} \cdot \bm{\sigma}} = T\sum_{p_0,\bm{p}}\Tr \left[G(p)(\bm{n} \cdot \bm{\sigma}) \right] =\sum_{\bm{p}} \left[n_{F+}- n_{F-} \right] \bm{n} \cdot \widehat{\bm{N}},
\end{split}
\end{equation}
where we substituted Eq.~\eqref{eq:GreensFunction}. The functions $n_{F\pm}= n_F[\varepsilon_\pm(\bm{p})-\mu]$ are Fermi distributions of the bands where $n_F(x) = [\exp(x/T)+1]^{-1}$ at chemical potential $\mu$ and temperature $T=0$. The factor $\left(n_{F+}- n_{F-}\right) $ restricts the momentum integration to the region between the two Fermi surfaces. The contribution from $\bm{j}^{(0)}$ is given in terms of the current-current correlation function,
\begin{equation}\label{eq:Kuboformula}
\begin{split}
 \Pi_{\alpha \beta}(\omega) = \lim_{\substack{i k_0 \rightarrow \\ \omega+ i\delta}}
-T\sum_{p_0, \bm{p}}  \Tr\left[ G(p^+) j^{(0)}_\alpha (\bm{p}) G(p^-)j^{(0)}_\beta (\bm{p}) \right] , 
 \end{split}
\end{equation}
where $p^\pm = p \pm k/2$, the four vector $k = (k_0,0)$ and $k_0$ is the bosonic Matsubara frequency. In the analytical continuation, $i k_0 \rightarrow \omega +i \delta$, where $\delta$ is a positive infinitesimal. Leaving the details of evaluating Eq.~(\ref{eq:Kuboformula}) to Appendix~\ref{Sec:Kubo-formula}, here we give the final result,
\begin{equation}\label{eq:Current-current-correlator-1}
\begin{split}
    \Pi_{\alpha \beta}(\omega) =&\sum_{\bm{p}} \left[ \frac{\p \bm{N}}{\p p_\alpha}\cdot\frac{\p \bm{N}}{\p p_\beta}-\left( \widehat{\bm{N}}\cdot\frac{\p \bm{N}}{\p p_\alpha}\right)\left( \widehat{\bm{N}}\cdot\frac{\p \bm{N}}{\p p_\beta}\right) \right] f_{-}(\omega, \bm{p}) \\
    & +i \sum_{\bm{p}}\widehat{\bm{N}}\cdot\left( \frac{\p \bm{N}}{\p p_\alpha}\times\frac{\p \bm{N}}{\p p_\beta} \right)f_{+}(\omega, \bm{p}),
\end{split}
\end{equation}
where we have introduced the functions 
\begin{equation}\label{eq:f-functions}
    f_{\pm}(\omega,\bm{p})=\left(n_{F+}- n_{F-}\right) \left(\frac{1}{\omega - 2|\bm{N}|+i\delta} \pm \frac{1}{\omega +2|\bm{N}|+i\delta} \right).
\end{equation}
The factor $\left(n_{F+}- n_{F-}\right) $ physically means that the photon excites an electron from the filled lower band to the unfilled upper band with band gap $2|\bm{N}|$; see Fig.~\ref{fig:Absorption}. The explicit expression of Eq.~\eqref{eq:Current-current-correlator-1} in terms of band parameters is given by Eq.~\eqref{eq:Current-Current-Correlator} in Appendix~\ref{Sec:Kubo-formula}.

In Eq.~(\ref{eq:Current-current-correlator-1}), the first term proportional to $f_-$ contributes to the symmetric component $\sigma^{\text{s}}(\omega)$ of the conductivity. As can be seen from Eq.~\eqref{eq:Conductivity}, $\sigma^{\text{s}}$ vanishes at $\omega\rightarrow 0$. This is due to neglecting impurity scattering which gives dissipation even at $\omega=0$. The second term provides the anti-symmetric part $\sigma^{\text{as}}(\omega)$ in Eq.~\eqref{eq:Conductivity} and is proportional to $\Omega(\bm{p}) $ given by Eq.~(\ref{eq:BerryCurvature-1}). At $\omega \rightarrow 0$, this term gives the anomalous Hall conductivity $\sigma_{\text{AH}}$. To see this, we substitute this term in Eq.~\eqref{eq:Current-current-correlator-1} into Eq.~\eqref{eq:Conductivity}, then take the limit $\omega\rightarrow 0$ to obtain the standard expression for $\sigma_{\text{AH}}$ of a two-band system~\cite{nagaosa2010anomalous}:
\begin{equation}\label{eq:AnomalousHallConductivity}
   \sigma^{\text{as}}_{xy} (0)=\sigma_{\text{AH}}= - e^2 \int\left(n_{F+} - n_{F-}\right)  \Omega(\bm{p})  \frac{d^2p}{(2\pi)^2}.
\end{equation}
Here we have used Eq.~(\ref{eq:BerryCurvature-1}) and the identity
\begin{equation}
    \lim_{\omega \rightarrow0} \frac{1}{\omega} \left(\frac{1}{\omega - 2|\bm{N}|+i\delta} + \frac{1}{\omega + 2|\bm{N}| +i\delta} \right)=-\frac{1}{2|\bm{N}|^2}.
\end{equation}
Note that for cases with $C=0$ but non-trivial $\Omega(\bm{p}) $, $\sigma_{\text{AH}}$ can still be non-zero due to the $(n_{F+} -n_{F-})$ factor in Eq.~\eqref{eq:AnomalousHallConductivity}. In what follows, we shall use $\sigma_{\text{AH}}$ to denote exclusively the anomalous Hall conductivity at zero frequency given by Eq.~\eqref{eq:AnomalousHallConductivity}. $\sigma_{xy}$ then refers to the $xy$ components of the conductivity tensor \eqref{eq:Conductivity} at arbitrary frequencies. In the cases where we compute $\sigma_{xy}(\omega)$, it includes only the anti-symmetric component $\sigma^{\text{as}}$~\eqref{eq:ASConductivity}.

Finally, note that Eq.~\eqref{eq:f-functions} can be rewritten using the identity
\begin{equation}
    \frac{1}{\omega \pm 2|\bm{N}|+i \delta} =  \text{P}\ \frac{1}{\omega \pm 2|\bm{N}|} - i\pi \delta(\omega \pm 2 |\bm{N}|);
\end{equation}
P denotes the principle value. The delta function corresponds to inter-band transitions since $2 |\bm{N}(\bm{p})|$ is the direct band gap at momentum $\bm{p}$. For frequencies below the minimum band gap $\Delta$ between the two Fermi surfaces, this term vanishes identically due to the $(n_{F+}- n_{F-})$ factor in Eq.~(\ref{eq:Current-current-correlator-1}), and there is no dissipation in the system in the limit of no impurities. This is due to the aforementioned impossibility of exciting an electron from the filled lower band to the unfilled upper band, as demonstrated schematically in Fig.~\ref{fig:Absorption}~(a). It can then be shown that $\sigma_{\alpha \beta}$ given by Eq.~\eqref{eq:Conductivity} satisfies the no-dissipation condition \eqref{eq:NoDissipation}.

In the rest of this section, we numerically compute (\ref{eq:Conductivity}) for the different magnetic-field cases considered in Sec.~\ref{Sec:Model}.

\subsection{No magnetic field}

We start from the simplest case in which $\bm{n} = \bm{e}_z$ and set the magnetic field to zero. The system then has $C_4T$ symmetry. Using Eq.~(\ref{eq:Conductivity-Symmetry}) and the fact that the anti-symmetry part in the conductivity in Eq.~\eqref{eq:Conductivity} changes sign under a $C_4$ rotation, we find that
\begin{equation}
\sigma_{xx} = \sigma_{yy}, \quad \text{and} \quad \sigma_{xy}=0.
\end{equation}
The longitudinal conductivity $\sigma_{xx}$ is plotted as a function of frequency $\omega$ in Fig.~\ref{fig:Absorption}~(b). In particular, the anomalous Hall response is absent despite the presence of magnetic order. This is because the system has a $4'$ symmetry, i.e. a four-fold rotation about the out-of-plane axis supplemented by a time reversal. This symmetry forbids the existence of a finite $g_z$ (and, hence, $\sigma^\text{as}_{xy}$) in Eq.~\eqref{eq:ASConductivity}.

\begin{figure}
    \centering
    \includegraphics[width=\linewidth]{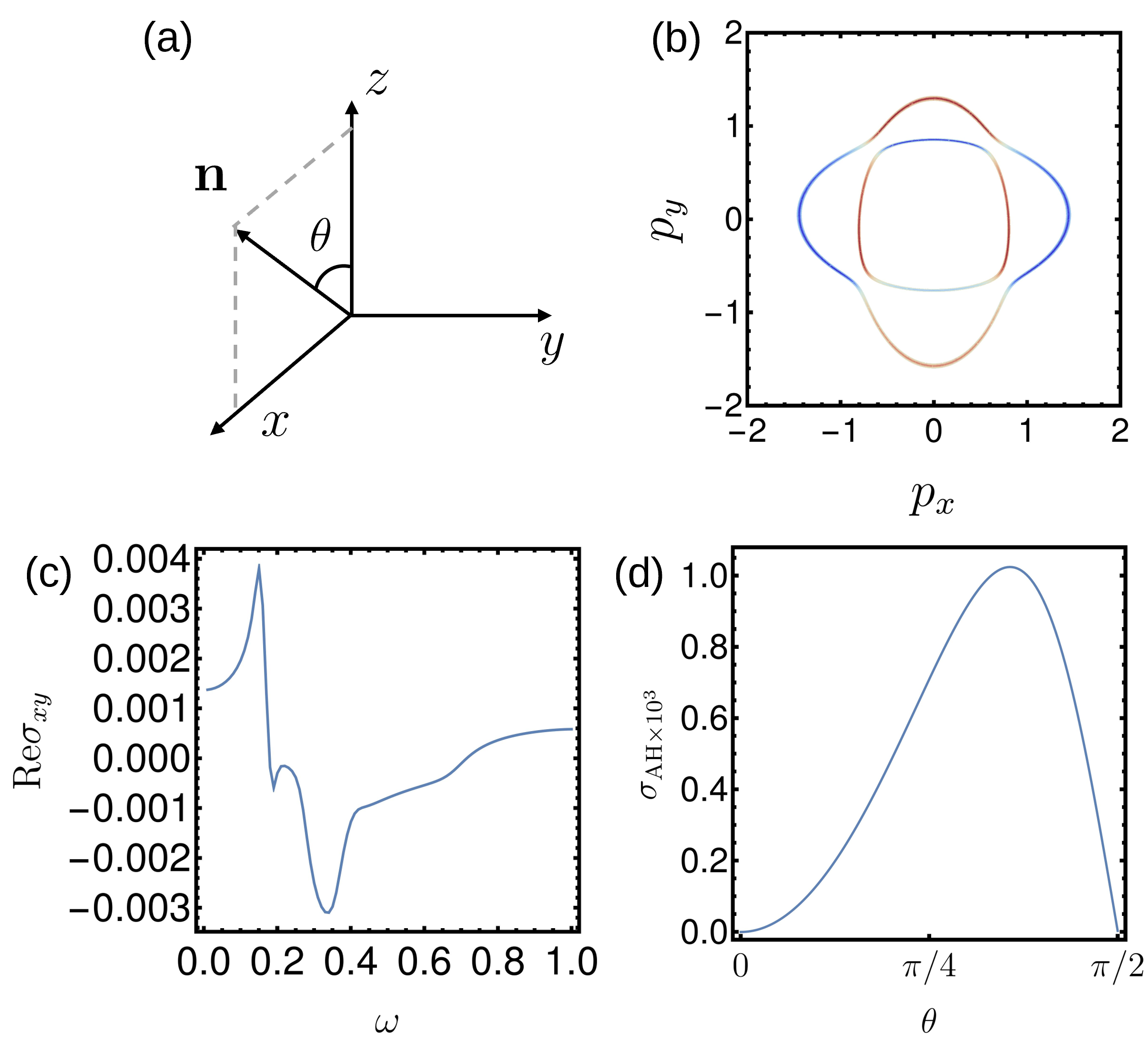}
    \caption{ Fermi surfaces and conductivitiy for $\bm{n}$ inside the $xz$-plane and $\mu=0.5t$ without magnetic field. (a) Schematic for the coordinate system, where $\theta$ is the azimuthal angle of $\bm{n}$ inside the $xz$-plane. (b) Fermi-surfaces at $\theta = \pi/4$.  (c) $\Re \sigma_{xy}$ given by Eq.~\eqref{eq:Conductivity} at $\theta = \pi/4$ as a function of frequency $\omega$. (d) Anomalous Hall conductivity $\sigma_{\text{AH}}$ in Eq.~(\ref{eq:AnomalousHallConductivity}) as a function of $\theta$.
    }
    \label{fig:Conductivity(H=0)}
\end{figure}

Tilting $\bm{n}$ away from the $z$-axis immediately generates a non-zero $\sigma^{\text{as}}$ as the $4'$ symmetry is now broken [see Fig.~\ref{fig:Conductivity(H=0)}~(a,b) for the setting and the Fermi surface plot], except for when $\bm{n}$ lies within the two mirror planes. In the latter case, there are $m'$ symmetries, i.e. mirror symmetries supplemented by time reversal, that forbid a finite $g_z$. The results are shown in Fig.~\ref{fig:Conductivity(H=0)}(c,d). For $\Re \sigma_{xy}$ we observe multiple peaks of the order $\alpha p_F$. This is due to the tilted $\bm{n}$, which distorts the four corners of the Fermi surface along the diagonals as shown in Fig.~\ref{fig:Conductivity(H=0)}~(b). The peaks then correspond to the direct inter-band transitions at these in-equivalent corners (for $\bm{n}=\bm{e}_z$ their positions coincide and the peaks cancel). For $\sigma_{\text{AH}}$ at $\omega=0$ in Eq.~\eqref{eq:AnomalousHallConductivity},  it is zero when $\bm{n} = \bm{e}_z$ as mentioned above. As $\bm{n}$ becomes in-plane, $\sigma_{\text{AH}}$ again vanishes. This is true more generally for the anti-symmetric component $\sigma^{\text{as}}(\omega)$ \eqref{eq:ASConductivity} at arbitrary frequency, because the corresponding contribution in Eq.~\eqref{eq:Current-current-correlator-1} is proportional to $\Omega(\bm{p}) $ which vanishes when $\widehat{\bm{N}}$ becomes a planar vector.

\subsection{Perpendicular magnetic field}

\begin{figure}
    \centering
    \includegraphics[width=0.6\linewidth]{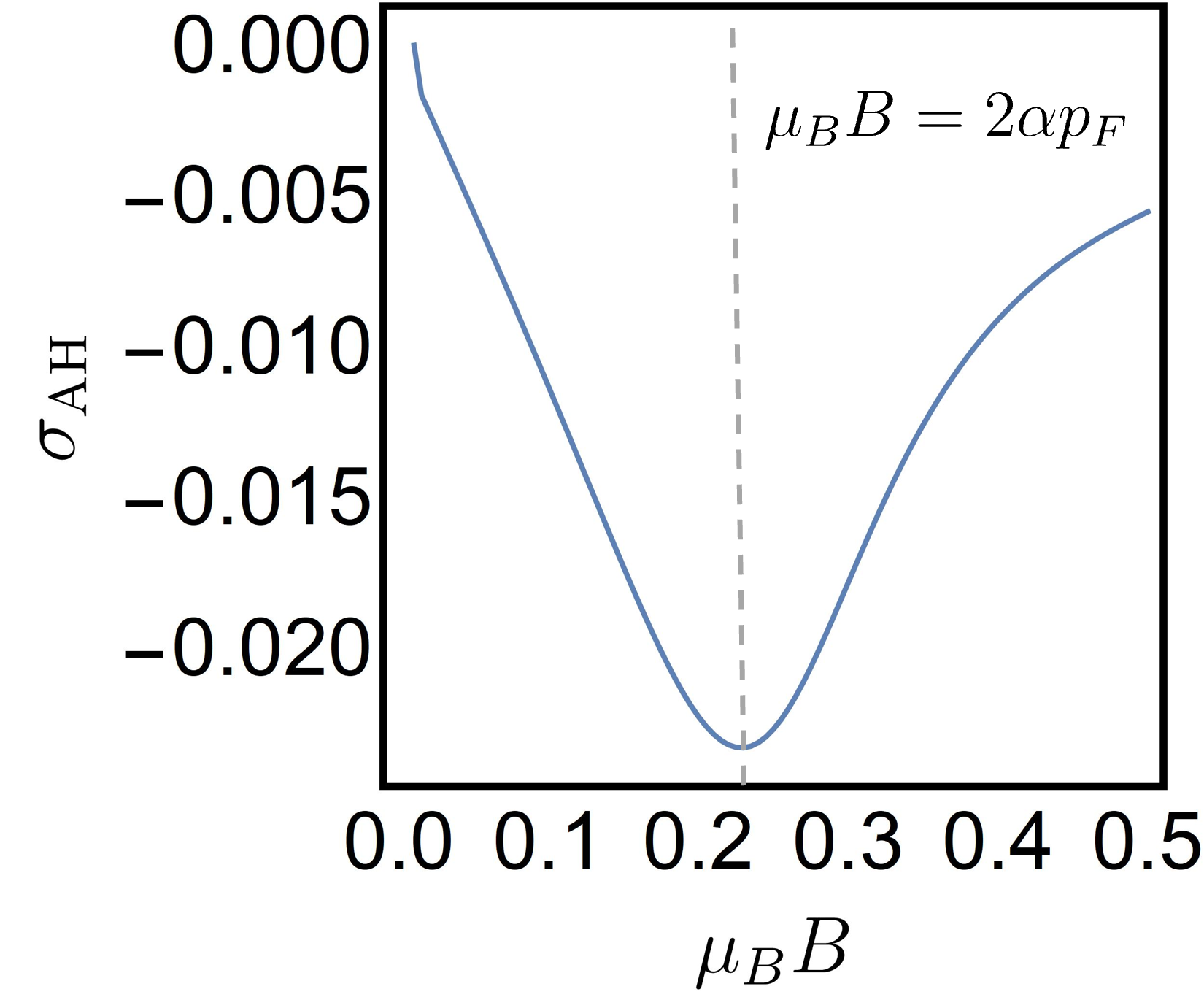}%
    \caption{Anomalous Hall conductivity $\sigma_{\text{AH}}$ in Eq.~(\ref{eq:AnomalousHallConductivity}) as a function of $\mu_B |\bm{B}|$  for $\bm{B}$ and $\bm{n}$ along $\bm{e}_z$. We have $ \sigma_{\text{AH}}\propto \mu_B |\bm{B}|$ up to  $\mu_B |\bm{B}|\approx 2\alpha p_F =0.2t$ in agreement with Eq.~(\ref{eq:Conductivity(Hperpendicular)}).
    }
    \label{fig:Conductivity(Hperpendicular)}
\end{figure}

We now return to the case of $\bm{n} = \bm{e}_z$. As discussed in Sec.~\ref{Sec:Model}, the dominant contribution to the Chern number comes from near the origin for small perpendicular field, whereas in Eq.~(\ref{eq:AnomalousHallConductivity}), only the region between the two Fermi surfaces contributes to $\sigma_{\text{AH}}$. Here the dominant contribution to $\sigma_{\text{AH}}$ comes from near the four intersectional lines, where the altermagnetic term vanishes. Near these points, $\Omega(\bm{p})  \sim \mu_B |\bm{B}|/(\alpha p_F^3)$ for $\mu_B |\bm{B}| \lesssim 2\alpha p_F$. The momentum area near the splitting in which this holds is $\delta p_\perp \delta p_\parallel$, where $\delta p_\parallel \sim \alpha /\beta_M$ is the momentum distance in the tangential direction along which the altermagnetic term dominates. As a result, $\sigma_{\text{AH}}$ is strongly suppressed by the small ratio $\mu_B |\bm{B}|/\mu$: 
\begin{equation}\label{eq:Conductivity(Hperpendicular)}
   \sigma_{\text{AH}}  \sim \Omega(\bm{p})  \delta p_\perp \delta p_\parallel \sim \left(\frac{\alpha p_F}{\mu}\right)\left(\frac{\mu_B |\bm{B}|}{\mu}\right).
\end{equation}
We plot $ \sigma_{\text{AH}}$ as a function of $\mu_B  |\bm{B}|$ in Fig.~\ref{fig:Conductivity(Hperpendicular)} which shows $ \sigma_{\text{AH}} \propto \mu_B  |\bm{B}|$ for $\mu_B |\bm{B}| \lesssim 2\alpha p_F$.

\subsection{In-plane magnetic field}

Despite that band topology and the Chern number $C$ can be tuned by an arbitrarily small in-plane magnetic field, for $\mu_B |\bm{B}| \ll \alpha p_F \ll \mu$, the effect on conductivity is small. This is because the change in the Berry curvature $\Omega(\bm{p}) $ is concentrated near the origin, but in Eq.~(\ref{eq:AnomalousHallConductivity}) the integration is taken between the two Fermi surfaces at momenta of order $p_F$. Therefore for the topological phase transitions to be observable in conductivity, it is necessary that the center of the Berry curvature quadrupole moments obeys $p_0 \sim p_F$. As follows from Eq.~(\ref{eq:MomentumOrigin}), this gives $\mu_B |\bm{B}|  \sim \alpha p_F$, namely when the Zeeman splitting becomes comparable to the Rashba splitting. 

\begin{figure}
    \centering
    \includegraphics[width=\linewidth]{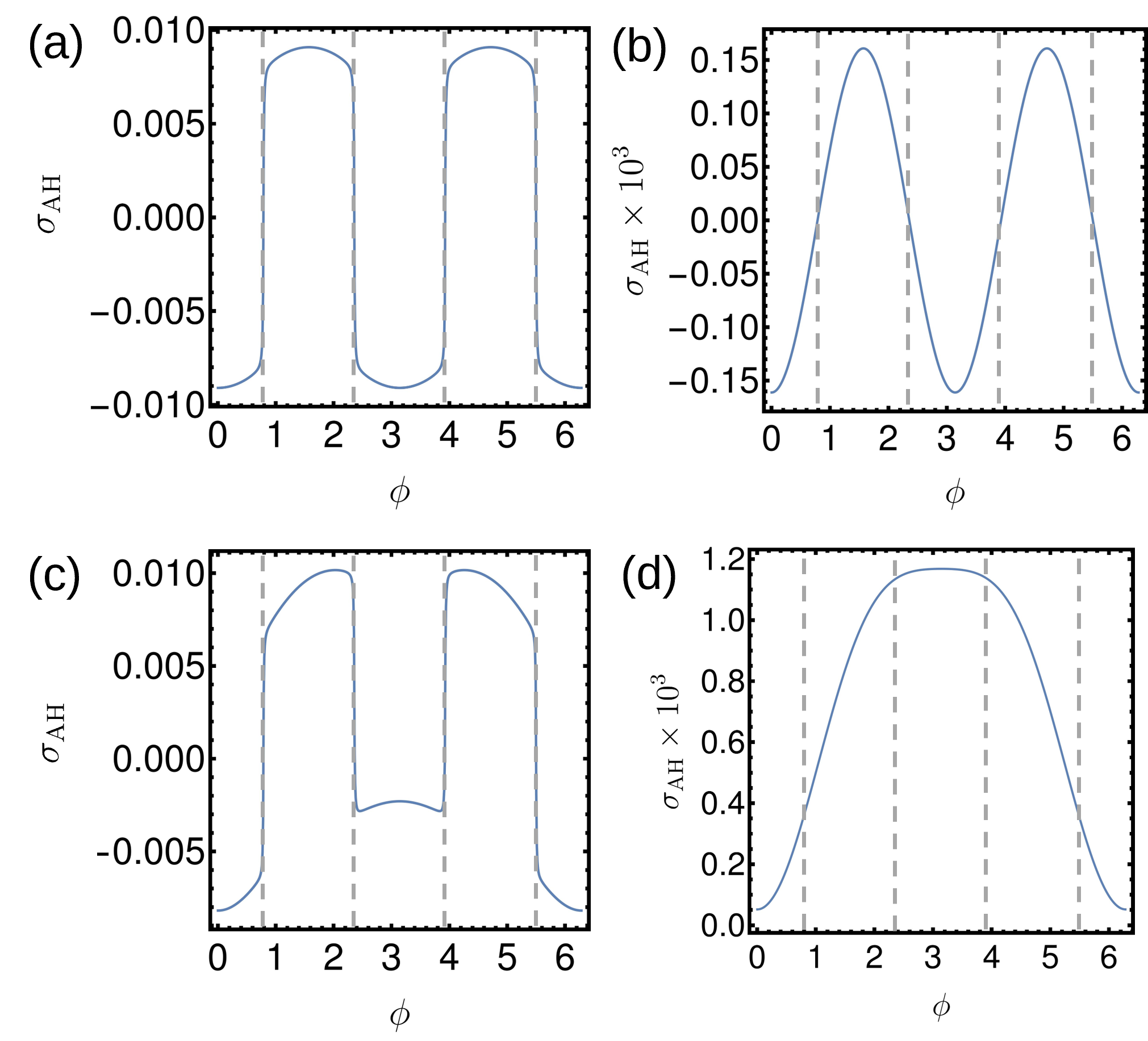}%
    \caption{ Anomalous Hall conductivity $\sigma_{\text{AH}}$ in Eq.~(\ref{eq:AnomalousHallConductivity}) as a function of in-plane magnetic field angle $\phi$.  Upper row: $\sigma_{\text{AH}}$ at $\bm{n}=(0,0,1)^\text{T}$ for (a) $\mu_B |\bm{B}| = 0.1$; and (b) $\mu_B|\bm{B}| = 0.01$. Lower row: $\sigma_{\text{AH}}$ at $\bm{n}=(1,0,1)^\text{T}/\sqrt{2}$ for (c) $\mu_B |\bm{B}| = 0.1$; and (d) $\mu_B |\bm{B}| = 0.01$. The vertical dashed lines are the in-plane topological transition points described in Sec.~\ref{sec:inplanefield}. Note the rescaled $\sigma_{\text{AH}}$ for $\mu_B |\bm{B}| = 0.01$, as $\sigma_{\text{AH}}$ decreases with decreasing field.
    }
    \label{fig:Conductivity(Hinplane)}
\end{figure}

In Figs.~\ref{fig:Conductivity(Hinplane)}(a,b), we illustrate the aforementioned relation between Zeeman and Rashba energies by plotting $\sigma_{\text{AH}}$ in Eq.~\eqref{eq:AnomalousHallConductivity} at $\bm{n}=\bm{e}_z$ as a function of the magnetic field's polar angle $\phi$. At gap closing angles, namely $\phi= \pi/4, 3\pi/4, 5\pi/4, 7\pi/4$, we see that $\sigma_{\text{AH}}$ changes discontinuously for $\mu_B |\bm{B}| = 0.1$ [see Fig.~\ref{fig:Conductivity(Hinplane)}(a)]. This is because $\bm{p}_0$ rotates with in-plane field and the gap closings occur at the four split corners of the Fermi surface. The discontinuous change is absent for a smaller Zeeman energy $\mu_B |\bm{B}| = 0.01$ [see Fig.~\ref{fig:Conductivity(Hinplane)}(b)]. For comparison, we also show the results for $\bm{n}=(1,0,1)/\sqrt{2}$ in Figs.~\ref{fig:Conductivity(Hinplane)}(c,d). These results show that, at $\bm{n}$ away from $\bm{e}_z$, signatures of topological phase transitions can still be seen by sharp jumps of $\sigma_{\text{AH}}$ [see Fig.~\ref{fig:Conductivity(Hinplane)}(c)] when $\mu_B |\bm{B}|  \sim \alpha p_F$. However at lower magnetic field, the features are smoothed out [see Fig.~\ref{fig:Conductivity(Hinplane)}(d)].

\section{Faraday effect}\label{Sec:FaradayEffect}

\begin{figure}
    \centering
    \includegraphics[width=\linewidth]{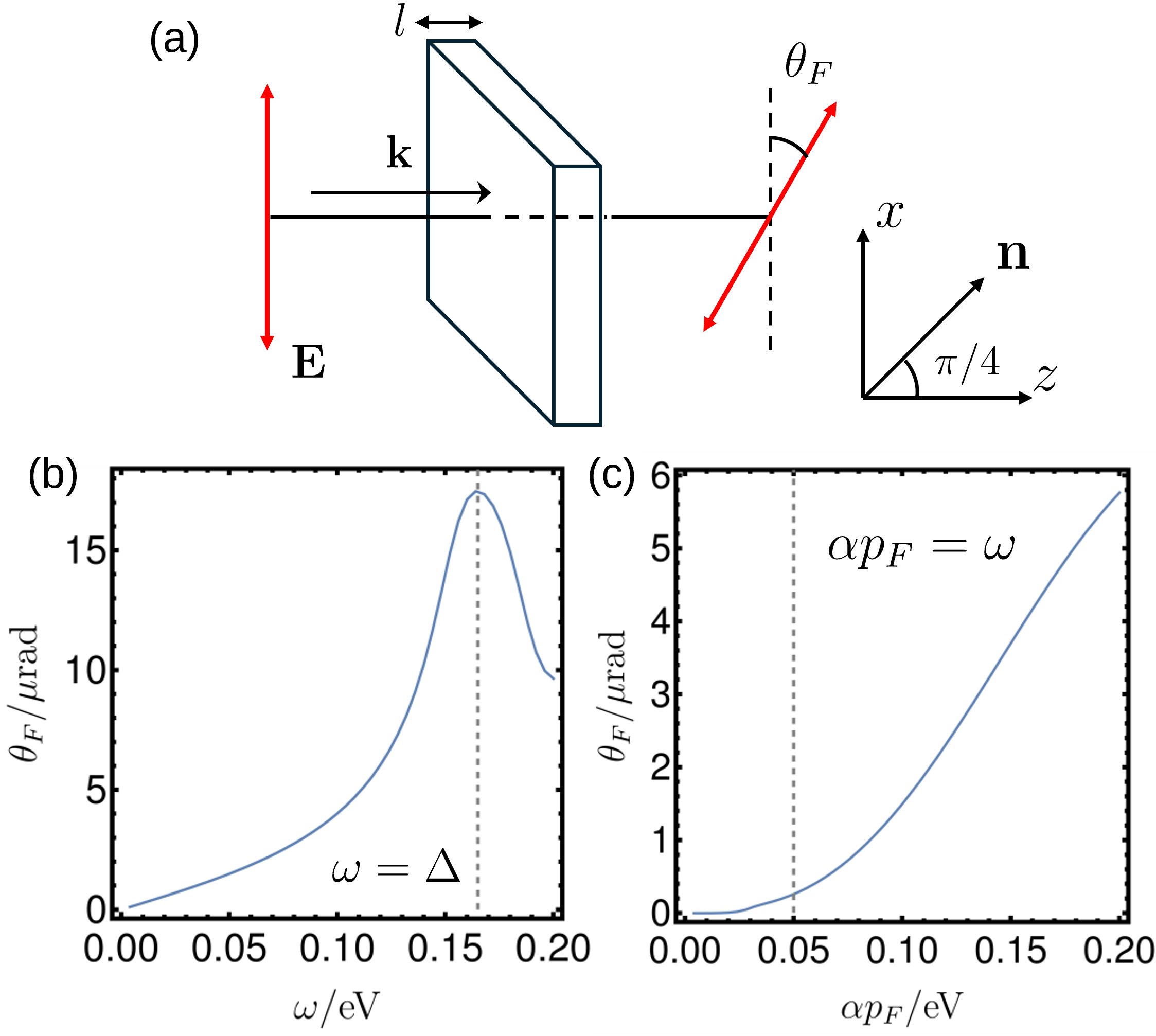}
    \caption{Faraday rotation with the N\'{e}el vector $\bm{n} = (1,0,1)^\text{T}/\sqrt{2}$. (a) Schematic plot for the Faraday effect, where linear polarized light is incident normally along the $z$-axis on the material with thickness $l$. The linear polarization is along the $x$-axis and is rotated by angle $\theta_F$ inside the $xy$-plane after passage of the material. (b) Faraday angle $\theta_F$ [see Eq.~\eqref{eq:FaradayAngle}] as a function of frequency $\omega$ for $\alpha = 0.1t, \beta_M =\mu = 0.5 t, m = 1/t$ and $\bm{B}=0$. The parameter $t=1$~eV. The gray line is the direct band-gap $\Delta$ above which Eq.~(\ref{eq:FaradayAngle}) becomes invalid. (c) $\theta_F$ as a function of $\alpha p_F$ for $\omega=0.05$~eV (gray-line). Eq.~(\ref{eq:FaradayAngle}) holds for $\omega \lesssim \alpha p_F \lesssim \mu = 0.5~$eV.
    }
    \label{fig:FaradayAngle}
\end{figure}

The breaking of time-reversal symmetry and the presence of a non-zero $\sigma^{\text{as}}$ also results in the Faraday effect: for frequencies below the band gap, the linear polarization of a normally incident light is rotated by the Faraday angle $\theta_F$ after passing through the material; see Fig.~\ref{fig:FaradayAngle}(a). We show that in the AM, $\theta_F$ is non-zero even without magnetic field.

To compute $\theta_F$, we follow the approach in Ref.~\cite{Landau8}. The main idea of the derivation is as follows. First we decompose the normally incident light into the two polarization eigenmodes $\bm{e}_\pm$ within the material with thickness $l$, each propagating with different wavenumbers $k_{\pm}$. The resulting polarization after the passage is then found by recombining the two eigenmodes again after propagating through the material. Since for two-dimensional systems $l \ll c/\omega$, we then take the limit $l \rightarrow 0$ in the final expressions.

We now decompose the incident light into the two eigenmodes given by Eq.~(\ref{eq:Eigenmodes}). These two eigenmodes are found in Appendix~\ref{Sec:FaradayAngle}. They are elliptically polarized inside the $xy$-plane:
\begin{equation}\label{eq:Eigenmodes}
    \bm{e}_+ = (\lambda_1, i\lambda_2,0 )^\text{T}, \ \bm{e}_- = (\lambda_2, -i\lambda_1 ,0)^\text{T}, \ \lambda_1^2+\lambda_2^2=1,
\end{equation}
with refractive indices $n_{\pm}$ respectively. The corresponding wave-numbers are $k_{\pm}=n_{\pm} \omega/c$. In Appendix~\ref{Sec:FaradayAngle}, we also give expressions for $n_{\pm}, \lambda_1, \lambda_2$ in terms of $\sigma_{\alpha \beta}$. 

In an anisotropic medium, the decomposition depends on the incident polarization $\bm{e}_0$. For simplicity we take $\bm{e}_0$ to be along the principle $x$-axis:
\begin{equation}
    \bm{e}_0 = (1,0,0)^\text{T} = \lambda_1 \bm{e}_+ + \lambda_2 \bm{e}_-.
\end{equation}
After propagating through the material of thickness $l$, the electric field has the form
\begin{equation}
    \bm{E} = E_0\left(\lambda_1 \bm{e}_+e^{ik_+l} + \lambda_2 \bm{e}_-e^{ik_-l}\right),
\end{equation}
which can be written in real form as
\begin{equation}\label{eq:ElectricField}
    E_x =  E_0\cos \omega t, \ E_y =  E_0 \rho \sin (\omega t+\varphi),
\end{equation}
where $\rho$ and $\varphi$ are given by
\begin{align}
 \rho &= \left| \frac{\lambda_1\lambda_2 \left(e^{i k_+ l } - e^{i k_- l } \right)}{\lambda_1^2 e^{i k_+ l } + \lambda_2^2 e^{i k_- l }} \right|, \label{eq:rho}\\
 \varphi &= \arg \left(\frac{ \lambda_1\lambda_2 \left(e^{i k_+ l } - e^{i k_- l } \right)}{ \lambda_1^2 e^{i k_+ l } + \lambda_2^2 e^{i k_- l }} \right). \label{eq:phi}
\end{align}
Equation~(\ref{eq:ElectricField}) satisfies
\begin{equation}
\frac{E_x^2}{\cos^2 \varphi} -\frac{2 \sin \varphi}{\rho \cos^2 \varphi}E_xE_y + \frac{E_y^2}{  \rho^2 \cos^2 \varphi } =E_0^2,
\end{equation}
corresponding to elliptically polarized light. Relative to the incident $x$-axis, its major axis is at an angle
\begin{equation}\label{eq:RotatedAngle}
    \theta_F=  \frac{1}{2} \arctan \left(\frac{2\rho \sin \varphi}{1-\rho^2} \right).
\end{equation}
and its eccentricity is given by
\begin{equation}\label{eq:eccentricity}
    \epsilon = \frac{2 \sqrt{\rho^4 + 2 \left( 2\sin^2 \varphi - 1 \right)\rho^2 + 1}}{\rho^2+1+ \sqrt{\rho^4 + 2 \left( 2\sin^2 \varphi - 1 \right)\rho^2 + 1}}.
\end{equation}
Physically, the incident linear polarization broadens into an ellipse and rotates with angle $\theta_F$ after passage through the material.

We now take the limit $k_{\pm }l\ll 1$ in Eqs.~\eqref{eq:rho} and \eqref{eq:phi} and obtain
\begin{align}\label{eq:RotatedAngle-1}
    &\rho = \left| \frac{ \lambda_1\lambda_2\left(k_+ -k_- \right)l}{ \lambda_1^2  + \lambda_2^2} \right|, \\
    &\varphi =\frac{\pi}{2}+ \arg \left(\frac{ \lambda_1\lambda_2\left(k_+ -k_- \right)l}{ \lambda_1^2  + \lambda_2^2} \right).
\end{align}
In a non-absorbing medium, $\varphi = \pi/2$ and $\epsilon=1$, that is to say the electromagnetic wave remains linearly polarized. The Faraday angle $\theta_F$ then becomes
\begin{equation}\label{eq:FaradayAngle}
\theta_F =  \arctan \rho.
\end{equation}
We check that in the isotropic limit and weak TRS breaking,  $n_\pm  \approx n_0 \mp |\sigma_{xy}|/(2\varepsilon_0 n_0)$ where $n_0$ is the isotropic refractive index at vanishing off-diagonal conductivity tensor $\sigma_{xy} = 0$. This gives the usual isotropic result $\theta_F= \sigma_{xy} /(2 \varepsilon_0 n_0 c)$, where $\sigma_{xy}$ is the conductivity \textit{per monolayer}~\cite{Landau8,Tse2011,Volkov1985}.

We now demonstrate the Faraday effect in altermagnets without magnetic field by numerically computing Eq.~(\ref{eq:FaradayAngle}) as a function of frequency for $\bm{n} = (1,0,1)^{\text{T}}/\sqrt{2}$ and $\bm{B}=0$: the principle axes of $\sigma_{\alpha\beta}$ are then along the $k_x$- and $k_y$-axes. To convert to physical units, we choose $t= 1~$eV (which is of the order of the band-width in RuO$_2$~\cite{Libor2022-1}), such that $\alpha p_F, \omega, \beta_M$ are in units of eV. The conductivity in Eq.~\eqref{eq:Conductivity} is then converted to units of $e^2/(\hbar l)$, where $l = 0.65~$nm is the thickness of an altermagnetic RuO$_2$ monolayer~\cite{Ko2018}. We find that $\theta_F\sim 10^{-5}~$rad for frequencies below the direct band-gap [see Fig.~\ref{fig:FaradayAngle}(b)]. For comparison, these values are one order of magnitude larger than the Kerr angles measured in cuprates $10^{-6}$~rad~\cite{Xia2008,He2011, Karapetyan2012,Kazuhiro2014}.

It is also shown in Appendix~\ref{Sec:FaradayAngle} that at small $\omega$:
\begin{equation}
     \theta_F  \propto \omega.
\end{equation}
In Fig.~\ref{fig:FaradayAngle}(b), this is observed for $\omega \lesssim \Delta$ the band gap. In Fig.~\ref{fig:FaradayAngle}(c) we also show $\theta_F$ as a function of $\alpha p_F$ for fixed frequency $\omega= 0.05~$eV. We see that the estimated $\theta_F$ values depend strongly on the strength of spin-orbit coupling.

\section{Conclusions} \label{Sec:Conclusion}

In this work, we investigated quantitatively the band topology and optical response in a two-dimensional d-wave altermagnetic metal with substrate-induced Rashba coupling. 

Using a low-energy Hamiltonian, we find that, with Rashba spin-orbit coupling and external magnetic field, the band exhibits non-trivial topology depending on the magnetic field directions. For given magnetic field strength but arbitrary directions, there are globally two topological phases characterized by Chern numbers $C= \pm1/2$. Changing field strength shifts the critical line. For in-plane magnetic field, there are four topological transitions at which the gap closes and $C$ changes by unity. These transitions correspond to magnetic field directions that lie within mirror planes of the system and do not depend on field strength.

It is also shown that the non-trivial band topology induces anomalous Hall conductivity $\sigma_{\text{AH}}$ that can be tuned by an external magnetic field. We demonstrate this explicitly by computing the conductivity tensor as functions of frequency and external field directions. It turns out that even if the band Chern number is zero, non-trivial Berry curvature $\Omega(\bm{p})$ can still result in non-zero $\sigma_{\text{AH}}$ by, for example tilting the N\'{e}el vector $\bm{n}$ that breaks the $4'$ symmetry. In cases with non-zero $C$, the correspondence between $C$ and $\sigma_{\text{AH}}$ is not straightforward, since only electrons lying between the two Fermi surfaces contribute to $\sigma_{\text{AH}}$. The effect of quantization of $C$ is the most pronounced when magnetic field is in-plane, and the Zeeman splitting is comparable to the Rashba splitting of the Fermi surfaces. In this case, the gap closings of the four topological transitions occur between the two Fermi surfaces, and $\sigma_{\text{AH}}$ undergoes discontinuous changes as the field rotates in plane. 

We also point out that, even though the anomalous Hall conductivity is not directly related to the Chern number for a metallic system, the non-trivial band topology is still expected to lead to chiral edge resonances. Despite their hybridization with gapless bulk modes and the absence of immunity against backscattering, the chiral edge resonance may still serve as a signature of the topological nature of metallic bands.

Finally we also showed, by an explicit numerical computation, that broken time-reversal-symmetry in our model results in a non-zero Faraday angle $\theta_F$.  Using parameters for RuO$_2$, $\theta_F$ is estimated to be of order $10^{-5}$ rad. Overall, the tuneability of the electric conductivity makes AM heterostructures a potentially useful platform in manufacturing non-reciprocal quantum devices without the need for external magnetic fields.

\section*{Acknowledgement:}
P. Rao thanks J. Habel, G. Sim and L. Classen for useful discussions.  J.K. acknowledges support from the Imperial-TUM flagship partnership. The research is part of the Munich Quantum Valley, which is supported by the Bavarian state government with funds from the Hightech Agenda Bayern Plus. A.M. was funded by the Deutsche Forschungsgemeinschaft (DFG, German Research Foundation) - Project No. 504261060.

\appendix

\section{Computation of Eq.~(\ref{eq:Current-current-correlator-1})}\label{Sec:Kubo-formula}
First we perform the trace over spin in Eq.~(\ref{eq:Kuboformula}) by substituting Eqs.~(\ref{eq:Hamiltonian-1}) and (\ref{eq:GreensFunction}), then using the identities
\begin{equation}
\begin{split}
    &\Tr (\sigma^a \sigma^b ) = 2 \delta_{ab}, \  \Tr (\sigma^a \sigma^b \sigma^c ) = 2 i \varepsilon_{abc}, \\
    &\Tr (\sigma^a \sigma^b \sigma^c \sigma^d ) = 2 (\delta_{ab} \delta_{cd} - \delta_{ac} \delta_{bd} + \delta_{ad} \delta_{bc}).
\end{split}
\end{equation}
The result is given by products of the electron Green's functions $G_{\pm}$ defined in Eq.~\eqref{eq:GreensFunction}. As will be shown below, only inter-band transition terms, that is, $G_\pm G_\mp$, remain after summing over $p_0$. Therefore, to simplify the expressions, we keep only these terms. Introducing the abbreviated notation $G_{+-} = G_+(p^-)$ etc., the result is
\begin{equation}\label{eq:Current-current-correlator-2}
\begin{split}
    \Pi_{\alpha \beta}(\omega) =&-T \sum_{p_0,\bm{p}} \bigg\{ \left[ \frac{\p \bm{N}}{\p p_\alpha}.\frac{\p \bm{N}}{\p p_\beta}-\left( \widehat{\bm{N}}.\frac{\p \bm{N}}{\p p_\alpha}\right)\left( \widehat{\bm{N}}.\frac{\p \bm{N}}{\p p_\beta}\right) \right] \\
    & \times(G_{++}G_{--}+ G_{-+}G_{+-}) +i \widehat{\bm{N}}.\left( \frac{\p \bm{N}}{\p p_\alpha}\times\frac{\p \bm{N}}{\p p_\beta} \right) \\
    & \times(G_{++}G_{--}- G_{-+}G_{+-}) \bigg\}.
\end{split}
\end{equation}
We now sum over the Matsubara frequency $p_0$ and perform the analytical continuation $ik_0 \rightarrow \omega +i\delta$ using the following identity
\begin{equation}\label{eq:Identities-1}
\begin{split}
  \lim_{i k_0 \rightarrow \omega+ i\delta} &T \sum_{p_0}   G_\alpha (ip_0 +ik_0,\bm{p}) G_\beta(ip_0,\bm{p})  \\
  =&\int_{-\infty}^{\infty}\frac{dz}{2\pi  } \tanh \frac{z}{2T} \big[\Im G^R_\alpha (z,\bm{p} ) G^A_\beta(z-\omega,\bm{p}) \\
  & + G^R_\alpha (z+\omega ,\bm{p})\Im G^R_\beta(z,\bm{p})  \big] \\
  =&-\left[n_{ F\alpha}(\bm{p})-  n_{F\beta}(\bm{p})\right] \frac{1}{\omega - [\varepsilon_\alpha(\bm{p})-\varepsilon_\beta(\bm{p})]+i\delta},
\end{split}
\end{equation}
where $G^{R,A}_\alpha (\omega,\bm{p}) = 1/[ \omega - \varepsilon_\alpha(\bm{p})\pm i\delta ]$ are the retarded and advanced Green's functions and $n_{ F\alpha}(\bm{p})= n_F[\varepsilon_\alpha(\bm{p})-\mu]$ is the Fermi distribution. Thus for finite $\omega$, only inter-band transition terms $\alpha \ne \beta$ remain, which physically correspond to direct band transitions induced by a photon with frequency $\omega$ and negligible momentum. We then have:
\begin{equation}\label{eq:Identities-2}
\begin{split}
   -T &\sum_{p_0} \left(G_{++}G_{--} \pm G_{-+}G_{+-} \right) \\
   &=\left(n_{F+} - n_{F-}\right) \left[\frac{1}{\omega - (\varepsilon_+-\varepsilon_-)+i\delta} \mp \frac{1}{\omega +\varepsilon_+-\varepsilon_- +i\delta} \right].
\end{split}
\end{equation}
Substituting Eq.~(\ref{eq:Identities-2}) in Eq.~(\ref{eq:Current-current-correlator-2}) and using the identity $\varepsilon_+ -\varepsilon_- = 2|\bm{N}|$ gives Eq.~(\ref{eq:Current-current-correlator-1}) in the main text. 

For reference, we provide the full expression for $\Pi_{\alpha \beta}(\omega)$ by substituting Eq.~(\ref{eq:Hamiltonian-1}) and the expression
\begin{align}
\frac{\p N_i}{\p p_\alpha} = \alpha \varepsilon_{i\alpha k}e_{z,k} + \beta_M n_i \tau^z_{\alpha \beta}p_\beta,
\end{align}
into Eq.~(\ref{eq:Current-current-correlator-1}), resulting in
\begin{equation}\label{eq:Current-Current-Correlator}
\begin{split}
    \Pi_{\alpha \beta}(\omega) &=\\
    &\sum_{\bm{p}} f_-(\omega,\bm{p}) \bigg\{\alpha^2 \left[ \delta_{\alpha \beta} - (\bm{e}_z \times \widehat{\bm{N}})_\alpha(\bm{e}_z \times \widehat{\bm{N}})_\beta\right] \\
    &+\beta_M^2 \tau^z_{\alpha \gamma} p_\gamma  \tau^z_{\beta \delta} p_\delta \left[1- (\widehat{\bm{N}}\cdot\bm{n})^2\right]  \\
    &+\alpha\beta_M \bigg(\tau^z_{\alpha \gamma} p_\gamma \left[(\bm{e}_z \times \bm{n})_\beta - (\bm{n}\cdot\widehat{\bm{N}})(\bm{e}_z \times \widehat{\bm{N}})_\beta\right]\\
    &+\tau^z_{\beta \gamma} p_\gamma [(\bm{e}_z \times \bm{n})_\alpha - (\bm{n}\cdot\widehat{\bm{N}})(\bm{e}_z \times \widehat{\bm{N}})_\alpha]\bigg) \bigg\} \\
    &+ i \sum_{\bm{p}} f_+(\omega,\bm{p})\bigg\{ \alpha^2 \widehat{N}_z \varepsilon_{\alpha \beta}+ \alpha \beta_M\bigg[\tau^z_{\alpha\gamma}p_\gamma \\
    &\left(\widehat{N}_\beta n_z - n_\beta\widehat{N}_z \right) + 
    \tau^z_{\beta\gamma}p_\gamma\left( n_\alpha\widehat{N}_z-  \widehat{N}_\alpha n_z  \right) \bigg] \bigg\}.
\end{split}
\end{equation}
The functions $f_\pm(\omega,\bm{p})$ are defined in Eq.~(\ref{eq:f-functions}). In evaluating the anti-symmetric term in Eq.~(\ref{eq:Current-Current-Correlator}), we have used
\begin{equation}
    \alpha^2 \varepsilon_{ijk} \widehat{N}_i \varepsilon_{j\alpha l}e_{z,l}  \varepsilon_{k\beta m}e_{z,m} =  \alpha^2 \varepsilon_{ijk} \widehat{N}_i \varepsilon_{j\alpha}  \varepsilon_{k\beta },
\end{equation}
since $\varepsilon_{ij z}= \varepsilon_{ij}$ is the two-dimensional anti-symmetric tensor. Using the identity $\varepsilon_{ij}\varepsilon_{kl}=\delta_{ik}\delta_{jl} - \delta_{il}\delta_{j k}$, this gives
\begin{equation}
    \alpha^2 \varepsilon_{ijk} \widehat{N}_i(\delta_{jk}\delta_{\alpha\beta} - \delta_{j\beta}\delta_{\alpha k}) =  \alpha^2 \varepsilon_{\alpha \beta} \widehat{N}_z,
\end{equation}
since $\alpha, \beta = x,y$ and $\varepsilon_{i\alpha \beta} = \varepsilon_{z\alpha \beta}$.

\section{Eigenmodes of light inside a dielectric medium}\label{Sec:FaradayAngle}

The eigenmodes of light propagation inside the material are determined by the dielectric tensor $\varepsilon_{\alpha\beta}$ [not to be confused with the eccentricity \eqref{eq:eccentricity}], which is related to the conductivity by the relation~\cite{Landau8}
\begin{equation}\label{eq:DielectricTensor}
    \varepsilon_{\alpha\beta} = \delta_{\alpha \beta} + \frac{ i}{\omega \varepsilon_0} \sigma_{\alpha \beta}.
\end{equation}
The vacuum permittivity is denoted by $\varepsilon_0$. The medium is assumed to be non-absorbing, $\omega < \Delta$, in which case it can be seen from Eq.~\eqref{eq:NoDissipation} that
\begin{equation}\label{eq:NoDissipationDielectric}
    \varepsilon_{\alpha\beta}(\omega) = \varepsilon_{\beta\alpha}^*(\omega),
\end{equation}
i.e., the symmetric and anti-symmetric components, $\varepsilon^\text{s}_{\alpha\beta}$ and $ \varepsilon^\text{as}_{\alpha\beta}$, are real and imaginary, respectively. The Onsager relation $\varepsilon_{\alpha \beta}(\omega; \bm{n}, \bm{B}) = \varepsilon_{\beta \alpha}(\omega; -\bm{n}, -\bm{B})$ then determines that $\varepsilon^\text{s}_{\alpha\beta}$ ($ \varepsilon^\text{as}_{\alpha\beta}$) is an even (odd) function of $\bm{n}$ and $\bm{B}$.

Our system corresponds to an anisotropic $\varepsilon_{\alpha\beta}$. The sample thickness $l \ll c/\omega$, which is always satisfied for thin films ($c$ is the speed of light). In the cases considered in Sec.~\ref{Sec:FaradayEffect}, the principle axes of $\varepsilon^\text{s}_{\alpha\beta}$ are always along the $x$- and $y$-axis. We fix the linear polarization of the normally incident wave along the $x$-axis.

We now find the eigenmodes. We assume the normal vector of the surface to be along one of the principle axes of $\varepsilon_{\alpha\beta}$. Then $\bm{E}$ is always transverse: $\bm{n} \cdot \bm{E}=0$ where $\bm{n} = \bm{k} c/\omega$ and $\bm{k}$ is taken along the $z$-axis. $\bm{E}$ then satisfies the Fresnel equation~\cite{Landau8}
\begin{equation}\label{eq:FresnelEquation}
 (n^2 \delta_{\alpha \beta} - \varepsilon_{\alpha \beta})E_\beta = 0.
\end{equation}
In the inverse dielectric tensor $\varepsilon_{\alpha\beta}$, we separate out the symmetric and anti-symmetric components,
\begin{equation}
     \varepsilon_{\alpha\beta} =  \varepsilon^{\text{s}}_{\alpha\beta}+ \varepsilon^{\text{as}}_{\alpha\beta}, 
\end{equation}
and choose the $x$-, $y$-axes to be the principle axes of $(\varepsilon^{\text{s}}_{\alpha\beta})$ with principle values $n_{0x}^2, n_{0y}^2$; $n_{0x}, n_{0y}$ are the refractive indices of the material. For the anti-symmetric part we write for later convenience $\varepsilon^{\text{as}}_{xy} = i G_z$. Equation~(\ref{eq:FresnelEquation}) then gives for the eigenmodes
\begin{equation}
n_{\pm}^2 = \frac{1}{2}\left(\varepsilon_{xx} + \varepsilon_{yy}\right) \pm \sqrt{\frac{1}{4}\left(\varepsilon_{xx} - \varepsilon_{yy}\right)^2 + G_z^2}.
\end{equation}
The corresponding elliptically polarized eigenvectors are given by Eq.~(\ref{eq:Eigenmodes}), where the coefficients $\lambda_1, \lambda_2$ satisfy
\begin{align}
\frac{\lambda_2}{\lambda_1}  = \frac{i}{G_z}  \left[\frac{1}{2}\left(n_{0x}^2 - n_{0y}^2\right) \mp \sqrt{\frac{1}{4}\left(n_{0x}^2 + n_{0y}^2\right)^2 + G_z^2} \right].
\end{align}
In a non-absorbing medium, it follows from Eq.~(\ref{eq:NoDissipationDielectric}) that $n_{0x}^2, n_{0y}^2$ and $G_z$ are real. Note that all intensive physical quantities above are defined \textit{per unit thickness}, whereas in two-dimensional materials they are \textit{per monolayer}.

Finally, we look at the limiting behavior of $\theta_F$ \eqref{eq:FaradayAngle} for $\omega \rightarrow 0$. Then $\varepsilon_{xx},\varepsilon_{yy} \sim 1/\omega^2\gg \varepsilon_{xy} \sim 1/\omega$ by substituting the diamagnetic term in Eq.~(\ref{eq:Conductivity}) into Eq.~(\ref{eq:DielectricTensor}). As a result, $\lambda_2/\lambda_1 \sim 1/\omega$ (remember $G_z \sim 1/\omega$) and $n_\pm \sim 1/\omega$. Equation~(\ref{eq:FaradayAngle}) then gives
\begin{equation}
    \tan \theta_F  \sim \frac{(n_+ - n_-) \omega}{(\lambda_1/\lambda_2) + (\lambda_2/\lambda_1)} \sim \omega \rightarrow 0.
\end{equation}

\nobalance 

%

\end{document}